\documentclass[acmlarge]{acmart}
\usepackage{fullpage}
\usepackage{url}
\usepackage{wrapfig}
\makeatletter
\def\@copyrightspace{\relax}
\makeatother
\usepackage{booktabs} 
\usepackage{mdframed}
\usepackage{framed}
\usepackage{hyphenat}
\usepackage{algpseudocode}
\usepackage{amsmath,epsfig}
\usepackage[boxed,ruled,vlined,linesnumbered]{algorithm2e}
\usepackage{amsthm}
\usepackage{setspace}
\usepackage{enumitem}
\usepackage{ragged2e}
\usepackage{multirow}
\usepackage{hhline}
\usepackage{wrapfig}
\usepackage{caption}
\usepackage{subcaption}
\usepackage{tabularx}
\usepackage{graphicx}
\usepackage{amsmath,epsfig}
\usepackage{amsthm}

\setcopyright{none}
\setcopyright{acmcopyright}
\setcopyright{acmlicensed}
\setcopyright{rightsretained}
\setcopyright{usgov}
\setcopyright{usgovmixed}
\setcopyright{cagov}
\setcopyright{cagovmixed}

\newcommand{\B}{\vspace*{-\smallskipamount}}
\newcommand{\BB}{\vspace*{-\medskipamount}}
\newcommand{\BBB}{\vspace*{-\bigskipamount}}

\usepackage{tikz}
\newcommand\encircle[1]{%
	\tikz[baseline=(X.base)]
	\node (X) [draw, shape=circle, inner sep=-1.5pt, fill=black, text=white] {\strut #1};}

\begin{document}

\title{\textsc{IoT Notary}: Attestable  Sensor Data Capture in IoT Environments}\titlenote{\textbf{A preliminary version of this work was accepted in IEEE NCA 2019 under the title ``\textsc{IoT Notary}: Sensor data attestation in smart environment.''\\ This version has been accepted in ACM Transactions on Internet Technology (TOIT), 2021.}}

\author{Nisha Panwar$^{1}$, Shantanu Sharma$^2$, Guoxi Wang$^3$, Sharad Mehrotra$^3$, Nalini Venkatasubramanian$^3$, Mamadou H. Diallo$^3$, and Ardalan Amiri Sani$^3$}
\affiliation{%
	\institution{$^1$Augusta University, USA. $^2$New Jersey Institute of Technology, USA. $^3$University of California, Irvine, USA.}}

\begin{abstract}
Contemporary IoT environments, such as smart buildings, require end-users to trust data-capturing rules published by the systems. There are several reasons why such a trust is misplaced --- IoT systems may violate the rules deliberately or IoT devices may transfer user data to a malicious third-party due to cyberattacks, leading to the loss of individuals' privacy or service integrity. To address such concerns, we propose \textsc{IoT Notary}, a framework to ensure trust in IoT systems and applications. \textsc{IoT Notary} provides secure log sealing on live sensor data to produce a verifiable `proof-of-integrity,' based on which a verifier can attest that captured sensor data adheres to the published data-capturing rules. \textsc{IoT Notary} is an integral part of TIPPERS, a smart space system that has been deployed at the University of California Irvine to provide various real-time location-based services on the campus. {\color{black} We present extensive experiments over realtime WiFi connectivity data to evaluate \textsc{IoT Notary}, and the results show that \textsc{IoT Notary} imposes nominal overheads. The secure logs only take 21\% more storage, while users can verify their one day's data in less than two seconds even using a resource-limited device.}
\end{abstract}


%

\maketitle


\section{Introduction}
\label{sec:introduction}
{\color{black}
Emerging IoT technologies~\cite{madakam2016,DBLP:journals/fgcs/GubbiBMP13} promise to bring revolutionary changes to domains including healthcare, wearable devices, transportation, smart buildings/homes, smart infrastructure, and emergency response, among others. IoT interconnects (a potentially large number of) commodity devices (\textit{e}.\textit{g}., sensors, actuators, controllers) into an integrated network that empowers systems with new capabilities and/or brings in transformational improvements to the existing system~\cite{DBLP:journals/fgcs/GubbiBMP13,DBLP:journals/iotj/KeohKT14}. In IoT systems, sensors are used for fine-grained monitoring of the evolving state of the infrastructure and the environment.
Our interest is in user-centric IoT spaces, wherein sensors of diverse types -- cameras, cell phones, WiFi access points, beacons, bodyworn sensors, occupancy sensors, temperature sensors, light sensors, and acoustic sensors -- are used to create awareness about subjects/end-users and their interactions among themselves and with the space.  While fine-grained continuous monitoring offers numerous benefits, it also raises several privacy and security concerns (\textit{e}.\textit{g}., smoking habits, gender, and religious belief).} To highlight the privacy concern, we first share our experience in building location-based services at UC Irvine using WiFi connectivity data.



\subsection*{Use-Case: University WiFi Data Collection}
In our ongoing project, entitled TIPPERS~\cite{DBLP:conf/percom/MehrotraKVR16}, we have developed a variety of location-based services based on a WiFi connectivity dataset. At UC Irvine, more than 2000 WiFi access-points and four WLAN controllers (managed by the university IT department) provide campus-wide wireless network coverage. Whenever a device connects to the campus WiFi network (through an access-point), the access-point generates a Simple Network Management Protocol (SNMP) trap for this association event. Each association event contains access-point-id, $s_i$, user device MAC address, $d_j$, and the time of the association, $t_k$. All SNMP traps $\langle s_i,d_j,t_k\rangle$ are sent to access-point's controllers in realtime. The access-point controller anonymizes device MAC addresses (to preserve the privacy of users in the campus).

TIPPERS collects WiFi connectivity data from one of the controllers that manage 490 access-points and receives $\langle s_i,d_j,t_k\rangle$ tuples for each connectivity event. However, before receiving any WiFi data, TIPPERS notifies all WiFi users about the data-capture rules by sending emails over a university mailing list. Subsequently, based on WiFi connectivity data $\langle s_i,d_j,t_k\rangle$, TIPPERS provides various realtime applications. Some of these services, \textit{e}.\textit{g}., computing occupancy levels of (regions in) buildings in the form of a live heatmap, require only anonymous data. Other services, \textit{e}.\textit{g}., providing location information (within buildings) or contextualized messaging (to provide messages to a user when he/she is in the vicinity of the desired location), require the user's original disambiguated data. To date, over one hundred users have registered into TIPPERS to utilize real-time services. A key requirement imposed by the university in sharing data with TIPPERS is that the system supports provable mechanisms to verify that individuals have been notified prior to their data (anonymized or not) being used for service provisioning. Also, an option for users to opt-out of sharing their WiFi connectivity data with TIPPERS must be supported. If users opt-out, the system must prove to the users that indeed their data was not shared with TIPPERS. TIPPERS uses immutable log-sealing to help all users to verify that the captured data is consistent with pre-notified data-capture rules.

Our experience in working with various groups in the campus is that (persistent) location data can be deemed quite sensitive by certain individuals with concerns about the spied upon by the administration or by others. Thus, mechanisms for notification of data-capture rules, secure log-sealing, and verification components made a sub-framework, entitled \textsc{IoT Notary}, which has become an integral part of TIPPERS. $\blacksquare$

\medskip
Data-capture concerns in IoT environments are similar to that in mobile computing, where mobile applications may have continuous access to resident sensors on mobile devices. In the mobile setting, data-capture rules and permissions~\cite{karmi} are used to control data access, \textit{i}.\textit{e}., which applications have access to which data generated at the mobile device (\textit{e}.\textit{g}., location and contact list) for which purpose and in which context. At the abstract level, a data-capture rule informs the user about the nature of personally identifying information that an application captures, \textit{i}.\textit{e}., for what {\em purpose} and in what {\em context}. However, in IoT settings, the data-capture framework differs from that in the mobile settings, in two important ways:

\begin{enumerate}[nolistsep,noitemsep,leftmargin=0.2in]
	\item Unlike the mobile setting, where applications can seek user's permission at the time of installation, in IoT settings, there are no obvious mechanisms/interfaces to seek users' preferences about the data being captured by sensors of the smart environment. Recent work~\cite{sup} has begun to explore mechanisms using which environments can broadcast their data-capture rules to users and seek their explicit permissions.
	\item Unlike the mobile setting, users cannot control sensors in IoT settings. While in mobile settings, a user can trust the device operating system not to violate the data-capture rules, in IoT settings, trust (in the environment controlling the sensors) may be misplaced. IoT systems may not be honest or may inadvertently capture sensor data, even if data-capture rules are not satisfied~\cite{madakam2016,aikins2016}.
\end{enumerate}

We focus on the above-mentioned second scenario and determine ways to provide trustworthy sensing in an untrusted IoT environment. Thus, the users can verify their data captured by the IoT environment based on pre-notified data-capture rules. Particularly, we deal with three sub-problems, namely \emph{secure notification} to the user about data-capture rules, \emph{secure (sensor data) log-sealing} to retain \emph{immutable} sensor data, as well as, data-capture rules, and \emph{remote attestation} to verify the sensor data against pre-notified data-capture rules by a user, without being heavily involved in the attestation process.

\subsection*{Our Contribution and Outline of the Paper}
In this paper, we provide the following:
\begin{itemize}[nolistsep,noitemsep,leftmargin=0.1in]
	\item A user-centric framework (\S\ref{sec:high_level_iot}) to ensure trustworthy data collection in untrusted IoT spaces, entitled \textsc{IoT Notary} that contains three entities (\S\ref{subsec:Entities}): {\color{black} (\textit{i}) \textit{infrastructure deployer} that installs sensors, (\textit{ii}) a \textit{service provider} (SP, \textit{e}.\textit{g}., TIPPERS) that establishes a list of data-capture rules that dictates the condition under which a sensor's data can/cannot be utilized to provide realtime services to the user, and (\textit{iii}) \textit{users} that use services provided by sensors, as well as, by SP, (if interested). }

	\item Two models to inform the user about the data-capture rules (\S\ref{subsec:notification phase}): notice-only model and notice-and-ACK model.
	
	\item A secure log-sealing mechanism (\S\ref{sec:Log Sealing}) implemented by secure hardware that cryptographically retains logs, data-capture rules, sensors' state, and contextual information to generate a \textit{proof-of-integrity} in an immutable fashion. 
	
	\item {\color{black} Optimized secure log-sealing mechanisms that regard the sensor state and user-device-id (\S\ref{subsubsec:Sealing Partial-No Sensor Data} and \S\ref{subsubsec:optimization}), implemented by secure hardware to reduce the size of the secure log.}
	
	\item A secure attestation mechanism (\S\ref{subsec:Attestation Phase}), mixed with SIGMA protocol~\cite{luks}, allowing a verifier (a user or a \emph{non-mandatory} auditor) to securely attest the sealed logs as per the data-capture rules.
	
	\item Implementation results of \textsc{IoT Notary} on the university live WiFi data in \S\ref{sec:Experimental Evaluation}. 
\end{itemize}

\subsection*{Advantages of \textsc{IoT Notary}} \textsc{IoT Notary} has the following distinct advantages:

\begin{enumerate}
  \item \textbf{Privacy-preserving system.} \textsc{IoT Notary} is the first privacy-preserving system for the enforcement of  IoT sensor data capturing policies and for the data attestation at the user to verify the presence/absence of their data against the data capturing policies. Moreover, during the policies' enforcement and attestation, any entity involved in \textsc{IoT Notary}  cannot learn anything about other other entities.

 \item \textbf{Minimal overhead in proof generation.}
 \textsc{IoT Notary} does not incur computational and space overhead in secure log generation. In particular, \textsc{IoT Notary} takes 310ms for creating the secure log over 37K rows and 21\% more space to store the log in a secure manner as compared to store the log in cleartext.

  \item \textbf{Scalable.}
  \textsc{IoT Notary} is a highly scalable system in terms of data verification. In particular, 
  a resource-constrained user (with a machine of 1-core 1GB RAM) can verify the presence or absence of their data over the data collected in a single day in most 30 seconds.


  \item \textbf{A deployed system.} At the University of California, Irvine, \textsc{IoT Notary} has been deployed and is in use to capture WiFi connectivity events. The real-world testbed has been established to provide a real environment where \textsc{IoT Notary}'s usability is tested and verified.
\end{enumerate}


\section{Modeling IoT Data Attestation}
\label{sec:Preliminaries}
This section presents the entities involved in \textsc{IoT Notary}, threat model, and the desired security properties.

\subsection{Entities}
\label{subsec:Entities}
\begin{figure}
	\centering
	\includegraphics[scale=0.36]{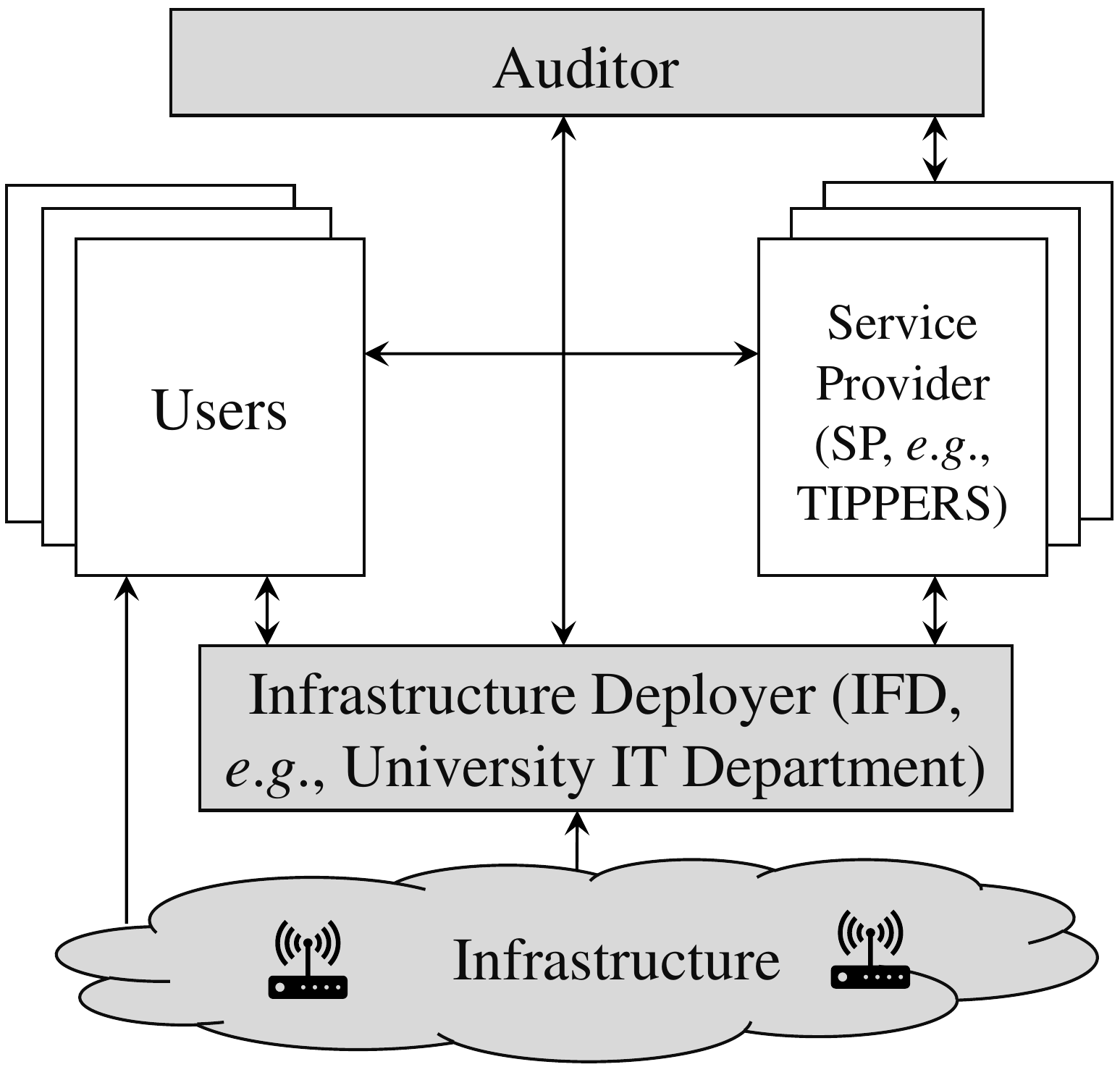}
	\caption{Entities in \textsc{IoT Notary}.}
	\label{fig:entity}
	\BBB
\end{figure}

Our model has the following entities, as shown in  Figure~\ref{fig:entity}:

\medskip
\noindent\textbf{Infrastructure Deployer (IFD).} IFD (which is the university IT department in our use-case; see \S\ref{sec:introduction}) deploys and owns a network of $p$ sensors devices (denoted by $s_1, s_2, \ldots, s_p$), which capture information related to users in a space. The sensor devices could be: (\textit{i}) dedicated sensing devices, \textit{e}.\textit{g}., energy meters and occupancy detectors, or (\textit{ii}) facility providing sensing devices, \textit{e}.\textit{g}., WiFi access-points and RFID readers. Our focus is on facility-providing sensing devices, especially WiFi access-points that also capture some user-related information in response to services. E.g., WiFi access-points capture the associated user-device-ids (MAC addresses), time of association, some other parameters (such as signal strength, signal-to-noise ratio); denoted by: $\langle d_i, s_j, t_k, \mathit{param} \rangle$, where $d_i$ is the $i^{\mathit{th}}$ user-device-id, $s_j$ is the $j^{\mathit{th}}$ sensor device, $t_k$ is $k^{\mathit{th}}$ time, and $\mathit{param}$ is other parameters (we do not deal with $\mathit{param}$ field and focus on only the first three fields). All sensor data is collected at a controller (server) owned by IFD. The controller may keep sensor data in cleartext or in encrypted form; however, it only sends encrypted sensor data to the service provider.

\medskip
\noindent\textbf{Service Providers (SP).} SP (which is TIPPERS in our use-case; see \S\ref{sec:introduction}) utilizes the sensor data of a given space to provide different \emph{services}, \textit{e}.\textit{g}., monitoring a location and tracking a person. SP receives encrypted sensor data from the controller.

\smallskip
\noindent\underline{\emph{Data-capture rules}}. SP establishes data-capture rules (denoted by a list $\mathcal{\mathit{DC}}$ having different rules $\mathit{dc}_1,\mathit{dc}_2, \ldots, \mathit{dc}_q$). Data-capture rules are conditions on device-ids, time, and space. Each data-capture rule has an associated \emph{validity} that indicates the time during which a rule is valid. Data-capture rules could be to capture user data by default (unless the user has explicitly opted out). Alternatively, default rules may be to opt-out, unless, users opt-in explicitly. Consider a default rule that individuals on the $6^{\mathit{th}}$ floor of the building will be monitored from 9pm to 9am. Such a rule has an associated condition on the time and the id of the sensor used to generate the data. Now, consider a rule corresponding to a user with a device $d_i$ opting-out of data capture based on the previously mentioned rule. Such an opt-out rule would have conditions on the user-id, as well as, on time and the sensor-id. For sensor data for which a default data-capture rule is opt-in, the captured data is forwarded to SP, if there do not exist any associated opt-out rules, whose associated conditions are satisfied by the sensor data. Likewise, for sensor data where the default is opt-out, the data is forwarded to SP only, if there exists an explicit opt-in condition. We refer to the sensor data to have a \emph{sensor state} ($s_i.\mathit{state}$ denotes the state of the sensor $s_i$) of 1 (or active), if the data can be forwarded to SP; otherwise, 0 (or passive). In the remaining paper, unless explicitly noted, opt-out is considered as the default rule, for simplicity of discussion.

Whenever SP creates a new data-capture rule, SP must send a \emph{notice message} to user devices about the current usage of sensor data (this phase is entitled \emph{notification phase}). SP uses Intel Software Guard eXtension (SGX)~\cite{sgx}, which works as a trusted agent of IFD, for securely storing sensor data corresponding to data-capture rules. SGX keeps all valid data-capture rules in the secure memory and only allows to keep such data that qualifies pre-notified valid data-capture rules; otherwise, it discards other sensor data. Further, SGX creates immutable and verifiable logs of the sensor data (this phase is entitled \emph{log-sealing phase}). The assumption of secure hardware at a machine is rational with the emerging system architectures, \textit{e}.\textit{g}., Intel machines are equipped with SGX~\cite{url1}. However, existing SGX architectures suffer from side-channel attacks, \textit{e}.\textit{g}., cache-line, branch shadow, page-fault attacks~\cite{DBLP:conf/ccs/WangCPZWBTG17}, which are outside the scope of this paper.

\medskip
\noindent\textbf{Users.} Let $d_1,d_2,\ldots,d_m$ be $m$ (user) devices carried by $u_1,u_2,\ldots,u_{m^{\prime}}$ users, where $m^{\prime}\leq m$. Using these devices, users enjoy services provided by SP. We define a term, entitled \emph{user-associated data}. Let $\langle d_i, s_j, t_k\rangle$ be a sensor reading. Let $d_i$ be the $i^{\mathit{th}}$ device-id owned by a user $u_i$. We refer to $\langle d_i, s_j, t_k\rangle$ as user-associated data with the user $u_i$. Users worry about their privacy, since SP may capture user data without informing them, or in violation of their preference (\textit{e}.\textit{g}., when the opt-out was a default rule or when a user opted-out from an opt-in default). Users may also require SP to prove service integrity by storing all sensor data associated with the user (when users have opted-in into services), while minimally being involved in the attestation process and storing records at their sides (this phase is entitled \emph{attestation phase}).

\medskip
\noindent\textbf{Auditor.} An auditor is a \emph{non-mandatory} trusted third-party that can (periodically) verify entire sensor data against data-capture rules. Note that a user can only verify his/her data, not the entire sensor data or sensor data related to other users, since it may reveal the privacy of other users.

\subsection{Threat Model}
\label{subsec:Threat Model}
We assume that SP and users may behave like adversaries. The adversarial SP may \emph{store} sensor data without informing data-capture rules to the user. The adversarial SP may \emph{tamper} with the sensor data by inserting, deleting, modifying, and truncating sensor readings and secured-logs in the database. By tampering with the sensor data, SP may \emph{simulate} the sealing function over the sensor data to produce secured-logs that are identical to real secured-logs. Thus, the adversary may hinder the attestation process and make it impossible to detect any tampering with the sensor data by the verifier (that may be an auditor or a user). Further, as mentioned before that SP utilizes sensor data to provide services to the user. However, an adversarial SP may provide \emph{false answers} in response to user queries. We assume that the adversarial SP cannot obtain the secret key of the enclave (by any means of side-channel attacks on SGX). Since we assumed that sensors are trusted and cannot be spoofed, we do not need to consider a case when sensors would collude with SP to fabricate the logs.

An adversarial user may \emph{repudiate} the reception of notice messages about data-capture rules. Further, an adversarial user may \emph{impersonate} a real user, and then, may retrieve the sensor data and secured-log during the verification phase. This way an adversarial user may reveal the privacy of the users by observing sensor data. Further, a user may infer the identity of other users associated with sensor data by potentially launching \emph{frequency-count attacks} (\textit{e}.\textit{g}., by determining which device-ids are prominent).

\subsection{Security Properties}
\label{subsec:Properties}
In the above-mentioned adversarial model, an adversary wishes
to learn the (entire/partial) data about the user, without notifying or by mis-notifying about data-capture rules, such that the user/auditor cannot detect any inconsistency between data-capture rules and stored sensor data at SP. Hence, a secure attestation algorithm must make it detectable, if the adversary stores sensor data in violation of the data-capture rules notified to the user. To achieve a secure attestation algorithm, we need to satisfy the following properties:

\medskip
\noindent\textbf{Authentication.} Authentication is required: (\textit{i}) between SP and users, during notification phase; thus, the user can detect a rogue SP, as well as, SP can detect rogue users, and (\textit{ii}) between SP and the verifier (auditor/user), before sending sensor data to the verifier to prevent any rogue verifier to obtain sensor data. Thus, authentication prevents threats such as impersonation and repudiation. Further, a periodic mutual authentication is required between IFD and SP, thereby discarding rogue sensor data by SP, as well as, preventing any rogue SP to obtain real sensor data.

\medskip
\noindent\textbf{Immutability and non-identical outputs.} We need to maintain the immutability of notice messages, sensor data, and the sealing function. Note that if the adversary can alter notice messages after transmission, it can do anything with the sensor data, in which case, sensor data may be completely stored or deleted without respecting notice messages. Further, if the adversary can alter the sealing function, the adversary can generate a proof-of-integrity, as desired, which makes flawless attestation impossible. The output of the sealing function should not be identical for each sensor reading to prevent an adversary to forge the sealing function (and to prevent the execution of a frequency-count attack by the user). Thus, immutability and non-identical outputs properties prevent threats, \textit{e}.\textit{g}., inserting, deleting, modifying, and truncating the sensor data, as well as, simulating the sealing function.

\medskip
\noindent\textbf{Minimality, non-refutability and privacy-preserving verification.} The verification method must find any misbehavior of SP, during storing sensor data inconsistent with pre-notified data-capture rules. However, if the verifiers wish to verify a subset of the sensor data, then they should not verify the entire sensor data. Thus, SP should send a minimal amount of sensor data to the verifier, enabling them to attest to what they wish to attest. Further, the verification method: (\textit{i}) cannot be refuted by SP, and (\textit{ii}) should not reveal any additional information to the user about all the other users during the verification process. These properties prevent SP to store only sensor data that is consistent with the data-capture rules notified to the user. Further, these properties preserve the privacy of other users during attestation and impose minimal work on the verifier.

\subsection{Assumptions}
\label{subsec:The Attestation Problem}
This section presents assumptions, we made, as follows:

\begin{enumerate}[noitemsep,nolistsep,leftmargin=0.2in]
	\item The sensor devices are assumed to be computationally inefficient to locally generate a verifiable log for the continuous data stream as per the data-capture rules.
	\item Sensor devices are tamper-proof, and they cannot be replicated/spoofed (\textit{i}.\textit{e}., two devices cannot have an identical id). In short, we assume a correct identification of sensors, before accepting any sensor-generated data at the controller at IFD, and it ensures that no rogue sensor device can generate the data on behalf of an authentic sensor. Further, we assume that an adversary cannot deduce any information from the dataflow between a sensor and the controller. (Recall that in our setting the university IT department collects the entire sensor data from their owned and deployed sensors, before sending it to TIPPERS.)
	\item We assume the existence of an authentication protocol between the controller and SP, so that SP receives sensor data only from authenticated and desired controller.
	\item The communication channels between SP and users, as well as, between SP and auditor are insecure. Thus, our solution incorporates an authenticated key exchange based on SIGMA protocol (which protects the sender's identity). When the verifier's identity is proved, the cryptographically sealed logs are sent to the verifier.
	
	\item By any side-channel attacks on SGX, one cannot tamper with SGX and retrieve the secret key of SGX. (Otherwise, the adversary can simulate the sealing process.)
	
	\item A hash function $H$ is known to the service provider, the auditor, and the user.
\end{enumerate}

{\color{black}
\medskip
\noindent\textbf{Definition: Attestation Protocol.} Now, we define an attestation protocol that consists of the following components:

\noindent $\bullet$ $\mathit{Setup}()$: Given a security parameter $1^k$, $\mathit{Setup}()$ produces a public key of the enclave ($\mathit{PK}_E$) and a corresponding private key ($\mathit{PR}_E$), used by the enclave to securely write sensor logs.

\noindent $\bullet$ $\mathit{Sealing}(\mathit{PR}_E, \langle d_i, s_j, s_j.\mathit{state}, t_k \rangle, \mathit{dc}_l)$: Given the key $\mathit{PR}_E$, $\mathit{Sealing}()$ (which executes inside the enclave) produces a verifiable \emph{proof-of-(log)-integrity} ($\mathcal{PI}$) and \emph{proof-of-integrity for user/service (query) verification} ($\mathcal{PU}$), based on the received sensor readings and the current data-capture-rule, $\mathit{dc}_l$.

\noindent $\bullet$ $\mathit{Verify}(\mathit{PK}_E, \langle \ast, s_j, s_j.\mathit{state}, t_k, \mathcal{PI}, \mathit{dc}_l\rangle, \mathit{Sealing}(\mathit{PR}_E, \\ \langle d_i, s_j, s_j.\mathit{state}, t_k \rangle, \mathit{dc}_l))$: 
Given the public key $\mathit{PK}_E$, sensor data, proof, and data-capture rule, $\mathit{Verify}()$ is executed at the verifier, where $\ast$ denotes the presence/absence of a user-device-id, based on $\mathit{dc}_l$. $\mathit{Verify}()$ produces 1, iff $\mathcal{PI}=\mathit{Verify}(\mathit{PK}_E, \langle \ast, s_j, s_j.\mathit{state}, t_k, \mathit{dc}_l\rangle,  \mathit{Sealing}(\mathit{PR}_E, \langle d_i, s_j, \\ s_j.\mathit{state}, t_k \rangle, \mathit{dc}_l))$; otherwise, 0. Similarly, $\mathit{Verify}()$ can attest $\mathcal{PU}$.

Note that the functions $\mathit{Sealing}()$ and $\mathit{Verify}()$ are known to the user, auditor, and SP. However, the private key $\mathit{PR}_E$ is only known to the enclave.
}

\B
\section{\textsc{IoT Notary}: Challenges and Approach}

This section provides challenges we faced during \textsc{IoT Notary} development and an overview of \textsc{IoT Notary}.

{\color{black}
\B
\subsection{Challenges and Solutions}
\label{subsec:Challenges and Solutions}
\B
We classify the problem of data attestation in IoT into three categories: (\textit{i}) secure notification, (\textit{ii}) log-sealing, and (\textit{iii}) log verification. Before going into details, we provide the challenges that we faced during the development and the way we overcome those challenges (Figure~\ref{fig:phases}), as follows:

\medskip
\noindent\textbf{C1. Secured notification for data-capture rules.} The declaration of data-capture rules requires a reliable and secure notification scheme, thereby users can verify the sender of notice messages. Trivially, this can be done through a unique key establishment between each user and SP. However, this incurs a major overhead at SP, as well as, SP can send different messages to different users.

\noindent\fbox{\begin{minipage}{3.3em}
		\noindent\textit{Solution.}
\end{minipage}}
To address the above challenge, we develop two solutions: One is based on a secure notifier that delivers a cryptographically encrypted notice message, which is signed by the enclave, to all the users (see \S\ref{subsec:notification phase}). Another solution uses an acknowledgment from the user and does not need any trusted notifier (see \S\ref{subsec:notification phase}). 


\medskip
\noindent\textbf{C2. Tamper-proof cryptographically secured-log sealing.} The verification process depends on immutable sensor data that is stored at SP. A trivial way is to store the entire data using a strong encryption technique, mixed with access-pattern hiding mechanisms. While it will prevent tampering with the data (except deletion), SP cannot develop real-time services on this data. Thus, the first challenge is how to store sensor data at SP according to data-capture rules; hence, the verifier can attest the data. The second challenge arises due to the computational cost at the verifier and communication cost between the verifier and SP to send the verifiable sensor data; \textit{e}.\textit{g}., if the verifier wishes to attest only one-day old sensor data over the sensor data of many years, then sending many years of sensor data to the verifier is impractical. Finally, the last challenge is how to securely store data when data-capture rules are set to be false (\textit{i}.\textit{e}., not to store data). In this case, not storing any data would not provide a way for verification, since SP may store data and can inform the verifier that there is no data as per the existing data-capture rules.

\noindent\fbox{\begin{minipage}{3.3em}
		\noindent\textit{Solution.}
\end{minipage}}
To address the first challenge, we develop a cryptographic sealing method based on hash-chains and XOR-linked-lists to ensure immutable sensor logs, after the sealed logs leave the enclave. Thus, any log addition/deletion/update is detectable (see \S\ref{subsubsec:Sealing Entire Sensor Data}). To address the second challenge, we execute sealing on small-sized chunks, which each maintains its hash-chain and XOR-links. The XOR-links ensure the log completeness, \textit{i}.\textit{e}., a chunk before and after the desired chunk has not been altered (see \S\ref{subsubsec:Sealing Entire Sensor Data}). To address the third challenge, we store the \emph{device state} of the first sensor reading for which the data-capture rule is set to false. Further, we discard all subsequent sensor readings, unless finding a sensor reading for which the data-capture rule is to \emph{store data} (\S\ref{subsubsec:Sealing Partial-No Sensor Data}).

\medskip
\noindent\textbf{C3. Privacy-preserving log verification.} In case of log-integrity verification, SP can provide the entire sensor data with a cryptographically sealed log to the trusted auditor. But, the challenge arises, if a user asks to verify her user-associated data/query results. Here, SP cannot provide the entire sensor data to the user, since it will reveal other users' privacy.

\noindent\fbox{\begin{minipage}{3.3em}
		\noindent\textit{Solution.}
\end{minipage}}
To address this challenge, we develop a technique to non-deterministically encrypt the user-id before cryptographically-sealing the sensor data. However, only non-deterministic encryption is also not enough to verify the log completeness, we compute XOR operation on each sensor reading, and then, create XOR-linked-list (see \S\ref{subsubsec:Sealing Data for Query Execution}).

\begin{figure*}[!t]
	
	\centering
	\includegraphics[scale=0.5]{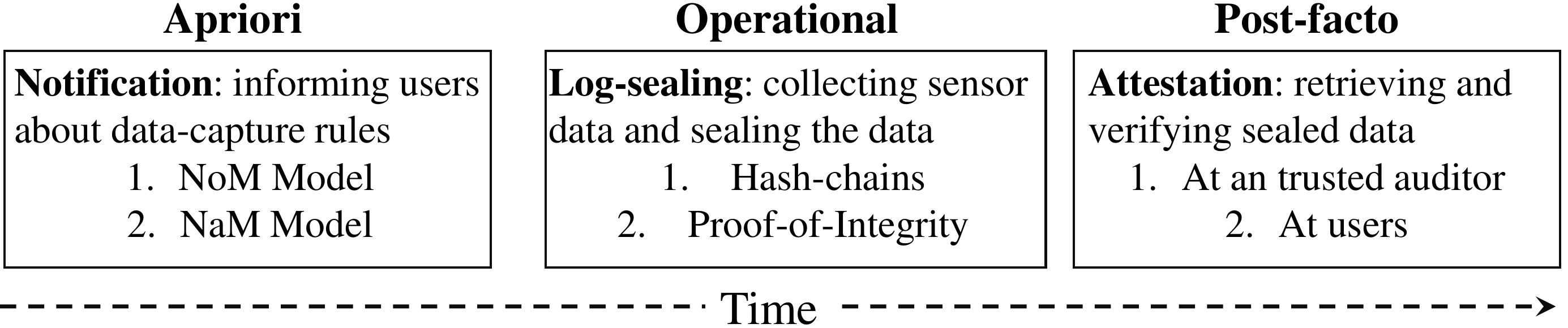}
	\caption{{\color{black} Phases in \textsc{IoT Notary}.}}
	\BB
	\label{fig:phases}
	\BB
\end{figure*}
}

\subsection{\textsc{IoT Notary}: An Overview}
\label{sec:high_level_iot}
\begin{figure*}[!t]
	\begin{center}
		\includegraphics[scale=0.5]{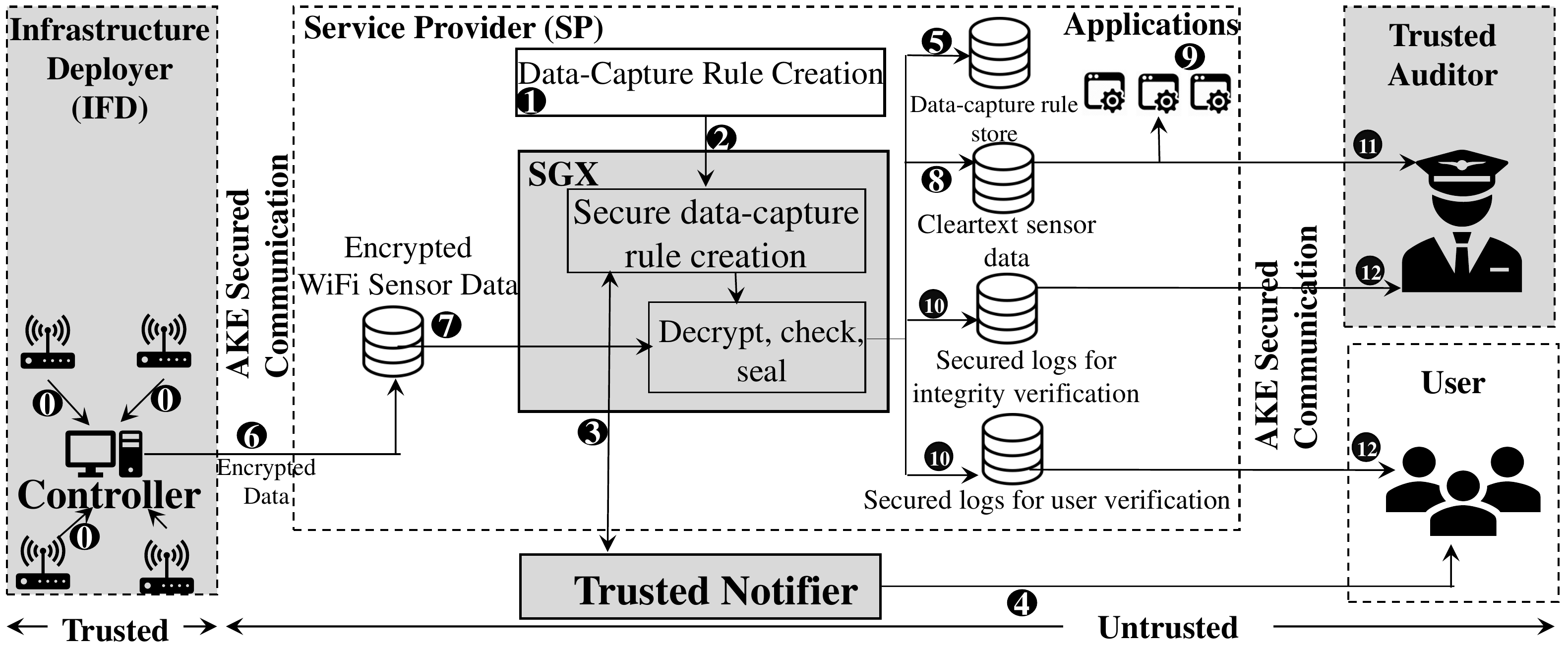}
	\end{center}
	\BB
	\caption{Dataflow and computation in the protocol. Trusted parts are shown in shaded box.}
	\BBB
	\label{fig:Dataflow and computation in the protocol}
\end{figure*}

This section presents an overview of the three phases and dataflow among different entities and devices, see Figure~\ref{fig:Dataflow and computation in the protocol}.

\medskip
\noindent\textbf{Notification phase: SP to Users messages.} This is the first phase that notifies users about data-capture rules for the IoT space using notice messages (in a verifiable manner for later stages). Such messages can be of two types: (\textit{i}) notice messages, and (\textit{ii}) notice-and-acknowledgment messages. SP establishes (the default) data-capture rules and informs trusted hardware (\encircle{1}). Trusted hardware securely stores data-capture rules (\encircle{2}, \encircle{5}) and informs the \emph{trusted notifier} (\encircle{3}) that transmits the message to all users (\encircle{4}). Only notice messages need a trusted notifier to transmit the message (see \S\ref{subsec:notification phase}).

\medskip
\noindent\textbf{Log-sealing phase: Sensor devices to SP messages.}
Each sensor sends data to the controller (\encircle{0}). The controller receives the correct data, generated by the actual sensor, as per our assumptions (and settings of the university IT department). The controller sends encrypted data to SP (\encircle{6}) that authenticates the controller using any existing authentication protocol, before accepting data. Trusted hardware (Intel SGX) at SP reads the encrypted data in the enclave (\encircle{7}).

\smallskip
\noindent\textit{\underline{Working of the enclave.}} The enclave decrypts the data and checks against the pre-notified data-capture rules. Recall that the decrypted data is of the format: $\langle d_i, s_j, t_k \rangle$, where $d_i$ is $i^{\mathit{th}}$ user-device-id, $s_j$ is the $j^{\mathit{th}}$ sensor device, and $t_k$ is $k^{\mathit{th}}$ time. After checking each sensor reading, the enclave adds a new field, entitled {\em sensor (device) states}. The sensor state of a senor $s_j$ is denoted by $s_j.\mathit{state}$, which can be \texttt{active} or \texttt{passive}, based on capturing user data. For example, $s_j.\mathit{state}$ = \texttt{active} or (\texttt{1}), if data captured by the sensor $s_j$ satisfies the data-capture rules; otherwise, $s_j.\mathit{state}$ = \texttt{passive} or (\texttt{0}). For all the sensors whose $\mathit{state} = 0$, the enclave deletes the data. Then, the enclave cryptographically seals sensor data, regardless of the sensor state, and provides cleartext sensor data of the format: $\langle d_i, s_j, s_j.\mathit{state}=1,t_k \rangle$ to SP (\encircle{8}) that provides services using this data (\encircle{9}). Note that the cryptographically sealed logs and cleartext sensor data are kept at untrusted storage of SP (\encircle{8}, \encircle{10}).

\medskip
\noindent\textbf{Verification phase: SP to verifier messages.} In our model, an auditor and a user can verify the sensor data. The auditor can verify the entire/partial sensor data against data-capture rules by asking SP to provide cleartext sensor data and cryptographically sealed logs (\encircle{8}, \encircle{10}). The users can also verify their own data against pre-notified messages or can verify the results of the services provided by SP using only cryptographically sealed logs (\encircle{12}). Note that using an underlying authentication technique (as per our assumptions), auditor/users and SP authenticate each other before transmitting data from SP to auditor/users.

\section{Attestation Protocol}
This section presents three phases of attestation protocol.
{\color{black} The important notations used in the presentation of the protocol are listed in Table~\ref{table:notations}.
}

\medskip
\noindent\textbf{Preliminary Setup Phase.} We assume a preliminary setup phase that distributes public keys ($\mathit{PK}$) and private keys ($\mathit{PR}$), as well as, registers user devices into the system. The trusted authority (which is the university IT department in our setup of TIPPERS) generates/renews/revokes keys used by the secure hardware enclave (denoted by $\langle \mathit{PK}_E, \mathit{PR}_E\rangle$) and the notifier (denoted by $\langle \mathit{PK}_N, \mathit{PR}_N\rangle$). The keys are provided to the enclave during the secure hardware registration process. Also, $\langle \mathit{PK}_{\mathit{di}}, \mathit{PR}_{\mathit{di}}\rangle$ denotes keys of the $i^{\mathit{th}}$ user device. \noindent\emph{Usages of keys}: The controller uses $\mathit{PK}_E$ to encrypt sensor readings before sending them to SP. $\mathit{PR}_E$ is also used by the enclave to write encrypted sensor logs and decrypt sensor readings. $\mathit{PK}_N$ is used during the notification phase by SGX to send an encrypted message to the notifier. User device keys are used during device registration, as given below.

We assume a registration process during which a user identifies herself to the underlying system. For instance, in a WiFi network, users are identified by their mobile devices, and the registration process consists of users providing the MAC addresses of their devices (and other personally identifiable information, \textit{e}.\textit{g}., email and a public key). During registration, users also specify their preferred modality through which the system can communicate with the user (\textit{e}.\textit{g}., email and/or push messages to the user device). Such communication is used during the notification phase.

\subsection{Notification Phase}
\label{subsec:notification phase}

The notification phase informs data-capture rules established by SP to the (registered) users by explicitly sending \emph{notice messages}. We consider two models for notification, differing based on acknowledgment from users.

In the \emph{notice-only model (NoM)}, SP informs users of data-capture rules, but users may not acknowledge receipt of the message. Such a model is used to implement policies, when data capture is mandatory, and the user cannot exercise control, over data capture. Since there is no acknowledgment, SP is only required to ensure that it sends a notice, but is not required to guarantee that the user received the notice. In contrast, a \emph{notice-and-ACK model (NaM)} is intended for discretionary data-capture rules that require explicit permission from users prior to data capture. Such rules may be associated, for instance, with fine-grained location services that require users' location. A user can choose not to let SP track his location, but will likely not be able to avail of some services.

Implementation of notification differs based on the model used. Interestingly, since NaM requires acknowledgment, the notification phase is easier as compared to NoM that uses a trusted notifier to deliver the message to users. Below we discuss the implementation of both models:

\noindent\emph{Notification implementation in NoM}. NoM assumes that, by default, data-capture rules are set not to retain any user data, unless SP, first, informs SGX about a data-capture rule, (\textit{i}.\textit{e}., SP cannot use the encrypted sensor data for building any application, see \encircle{9} in Figure~\ref{fig:Dataflow and computation in the protocol}). When SP creates a new data-capture rule, SP must inform SGX. Then, the enclave encrypts the data-capture rule using the public key (\textit{i}.\textit{e}., $\mathit{PK}_N$) of the notifier and informs the trusted notifier (via SP) about the encrypted data-capture rule by writing it outside of the enclave (in our user-case \S\ref{sec:introduction}, the university IT department works as a trusted notifier). Data-capture rules are maintained by SP on stable storage, which is read by SGX into the enclave to check, if the sensor data should be forwarded to SP. SGX can retain a cache of rules in the enclave, if such rules are still valid (and hence used for enforcement).\footnote{{\scriptsize Since the enclave has limited memory, the enclave cannot retain all the valid and non-valid data-capture rules after a certain size. Thus, the enclave writes all the non-valid data-capture rules on the disk after computing a secured hash digest over all the rules. Taking a hash over the rules is required to maintain the integrity of all the rules, since any rule written outside of the enclave can be tampered by SP. Recall that altering a rule will make it impossible to verify partial/entire sensor data.}} Finally, the trusted notifier acknowledges SP about receiving the encrypted data-capture rule, and then, informs users of the encrypted data-capture rule via signed notice messages. On receiving the notice message, the users may decrypt it and obtain the data-capture rule.


To see the role of \emph{trusted hardware} above, suppose that SP was responsible for informing users about data-capture rules directly. Data-capture rules are also required by SGX during log-sealing (\textsc{Phase} 2). An adversarial SP may inform SGX, not to users, or may inform non-identical rules to users and to SGX. Hence, SP cannot inform the rule to users directly.

To see the role of the \emph{trusted notifier} above, suppose that SP can directly inform users about encrypted data-capture rules obtained from SGX. An adversarial SP may not deliver the data-capture rule to all or some of the users; thus, an encrypted data-capture rule is not helpful. Hence, a trusted notifier ensures that the notice message is sent to all the registered users. Note that the trusted notifier might be a trusted website that lists all the data capture rules which users can access.

\noindent\emph{Implementation of notification in NaM}. Unlike NoM, the notification phase of NaM does not require the trusted notifier. In NaM, by default, SP cannot utilize all those sensor readings having device-ids for which the users have not acknowledged. Likewise NoM, in NaM, SP informs data-capture rules to SGX that encrypts the rule and writes outside of the enclave. The encrypted rules are delivered by SP to users, unlike NoM. On receiving the message, a user may securely acknowledge the enclave about her consent. The enclave retains all those device-ids that acknowledge the notice message for the log-sealing phase and considers those device-ids during the log-sealing phase to retain their data while discarding data of others.

\subsection{Log Sealing Phase}
\label{sec:Log Sealing}
The second phase consists of cryptographically sealing the sensor data for future verification against pre-notified data-capture rules. The sensor data is sealed into secured logs using authenticated data structures, \textit{e}.\textit{g}., hash-chains and XOR-linked lists (as shown in Figures~\ref{fig:sealing function execution},~\ref{fig:eoc}), by the sealing function, $\mathit{Sealing}(\mathit{PR}_E, \langle d_i, s_j, s_j.\mathit{state}, t_k\rangle)$, executed in the enclave at SP. Let us explain log-sealing in the context of WiFi connectivity data. The enclave reads the encrypted sensor data (\encircle{7} in Figure~\ref{fig:Dataflow and computation in the protocol}) and executes the three steps: (\textit{i}) decrypts the data, (\textit{ii}) checks the data against pre-notified valid data-capture rules, and (\textit{iii}) cryptographically seals the data and store \emph{appropriate secured logs}.

Below we explain our log sealing approach. To simplify the discussion, we first consider the case when all the sensor data satisfies some data-capture rule (\textit{i}.\textit{e}., the state of all the sensor data is one), and hence, data is forwarded to and stored at SP~\S\ref{subsubsec:Sealing Entire Sensor Data}. Then, we adapt the protocol to deal with the case when some sensor data satisfies some data-capture rule (\textit{i}.\textit{e}., the state of some sensor data is one, and hence, data is forwarded to and stored at SP), while remaining sensor data does not satisfy any rule (\textit{i}.\textit{e}., the state of the remaining sensor data is zero, and hence, data is forwarded to SP)~\ref{subsubsec:Sealing Partial-No Sensor Data}.

{\color{black}
\begin{table}[htbp]
{\color{black}
\begin{center}
\begin{tabular}{r c p{10cm} }
\toprule
\multicolumn{3}{c}{\underline{Sensor data after checking data-capture rule in the enclave}}\\
\multicolumn{3}{c}{}\\
$d_{i}$ & $\triangleq$ & $i^{\mathit{th}}$ user-device-id\\

$t_{k}$ & $\triangleq$ & $k^{\mathit{th}}$ time\\

$s_{j}$ & $\triangleq$ & $j^{\mathit{th}}$ sensor device\\

$s_{j}.state$ & $=$ & \(\left\{\begin{array}{rl}
1,  & \text{if the sensor $s_j$'s state is active; the data can be forwarded to SP} \\
0,  & \text{otherwise} \end{array} \right.\)\\

\multicolumn{3}{c}{}\\
\multicolumn{3}{c}{\underline{Data in log sealing phase}}\\
\multicolumn{3}{c}{}\\

$\mathcal{C}_x$ & $\triangleq$ & $x^{\mathit{th}}$ data chunk \\

$g^b$ & $\triangleq$ & A random string for chunk $\mathcal{C}_x$\\
$S^x_{eoc}$ & $\triangleq$ & XORed value of random string $g^a$,$g^b$,$g^c$ from chunk $\mathcal{C}_v$, $\mathcal{C}_x$, $\mathcal{C}_y$\\

$h_i^x$ & $\triangleq$ & The hash digest of $i^{\mathit{th}}$ sensor reading in chunk $\mathcal{C}_x$ (for log integrity)\\

$H$ & $\triangleq$ & A hash function known to the service provider, the auditor, and the user \\

$hu_i^x$ & $\triangleq$ & The hash digest of $i^{\mathit{th}}$ sensor reading in chunk $\mathcal{C}_x$ (for user  verification)\\

$\mathcal{PI}$ & $\triangleq$ & Proof-of-(log)-integrity of a chunk, used by the trusted auditor \\

$\mathcal{PU}$ & $\triangleq$ & Proof-of-integrity of a chunk, for the user/service (query) verification \\
\bottomrule

\end{tabular}
\end{center}
}\caption{{\color{black}Table of Notation for the Attestation Protocol}}
\label{table:notations}
\end{table}
}

\subsubsection{\textbf{{\underline{Sealing Entire Sensor Data}}}}
\label{subsubsec:Sealing Entire Sensor Data}
Informally, the sealing function executes a hash function on each sensor reading (or value), whose output is used to create a chain of hash digests. At the end of the sensor readings/values, the sealing function generates an authenticated proof-of-integrity by mixing a computationally-hard secure string. For example, consider four values: $v_1$, $v_2$, $v_3$, and $v_4$. The sealing function works as follows:

\begin{center}
	\begin{tabular}{|l|l|l|l|l|}\hline
		Value & Hash output\\\hline
		$v_1$ & $h_1\leftarrow H(v_1||H(0))$\\\hline
		$v_2$ & $h_2\leftarrow H(v_2||h_1)$\\\hline
		$v_3$ & $h_3\leftarrow H(v_3||h_2)$\\\hline
		$v_4$ & $h_4\leftarrow H(v_4||h_3)$\\\hline
		Proof-of-integrity & $\langle \mathit{sign}_{\mathit{PR}_E}(\mathit{SS} \oplus h_4)\rangle $ \\\hline
	\end{tabular}
\end{center}

In this example, all the values are hashed while including the hash digest of the previous value, except the first value. Note that only the first value is hashed with the hash digest of zero. In the end, the proof-of-integrity is prepared by signing XORed-value of the hash digest of the last value and a secret random string, $\mathit{SS}$. Note that the secret random string is generated for each chunk in a specific manner, which will be clear in the detailed description of the protocol; please see below.

The sealing operation consists of the following three phases: (\textit{i}) chunk creation, (\textit{ii}) hash-chain creation, and (\textit{iii}) proof-of-integrity creation; described below.

\begin{figure*}[!t]
	\BBB
	\begin{center}
		\includegraphics[scale=0.5]{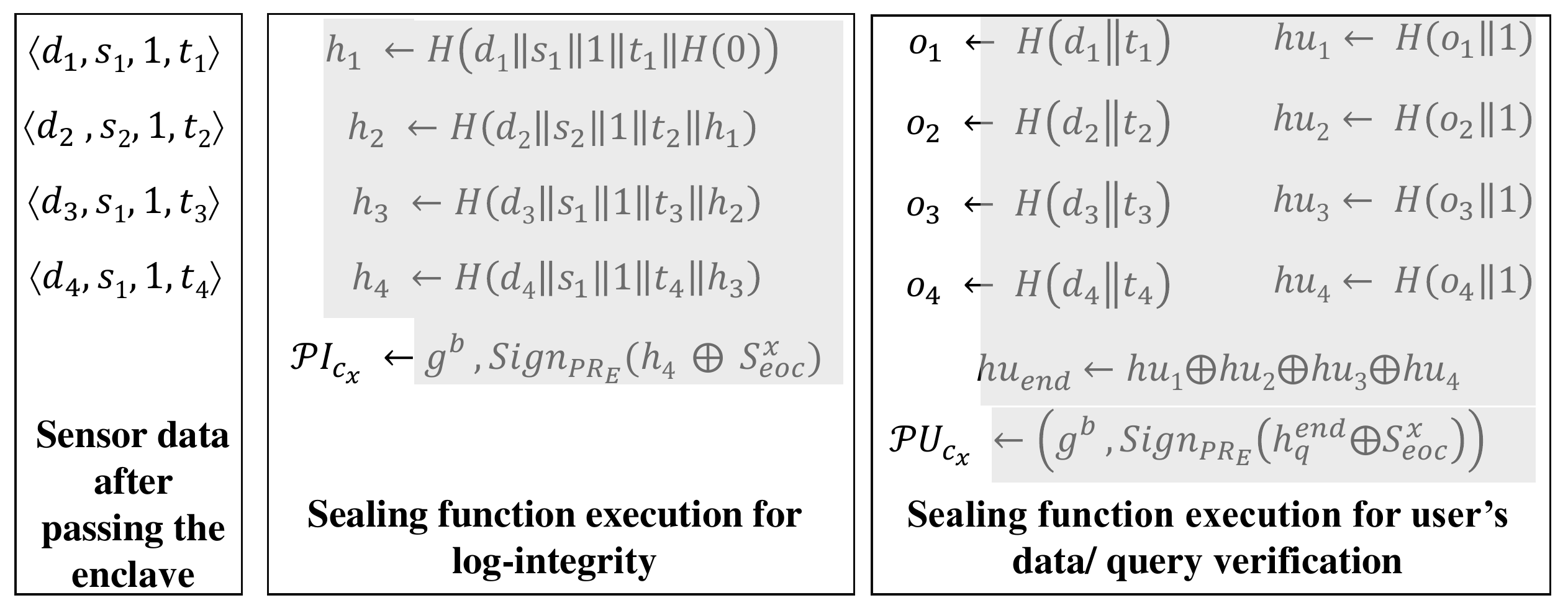}
	\end{center}
	\B
	\caption{Cryptographically sealing procedure executed on a chunk, $\mathcal{C}_x$.  Gray-shaded data is not stored on the disk. White-shaded data is stored on the disk and accessible by SP. The figure shows proof-of-integrity only for one chunk, $\mathcal{C}_x$; hence, some notations are abusively used. Note that we used $h_i$ to denote a hash digest computed for verifying the entire sensor data, while $\mathit{hu}_i$ denotes a hash digest computed for verifying the user data or query results.}
	\BB
	\label{fig:sealing function execution}
\end{figure*}

\medskip
\noindent\textbf{\textsc{Phase} 1: Chunk creation.} The first phase of the sealing operation finds an appropriate size of a chunk (to speed up the attestation process). Note that the incoming encrypted sensor data may be large, and it may create problems during verification, due to increased communication between SP and the verifier. Also, the verifier needs to verify the entire data, which has been collected over a large period of time (\textit{e}.\textit{g}., months/years). Further, creating cryptographic sealing over the entire sensor data may also degrade the performance of $\mathit{Sealing}()$ function, due to the limited size of the SGX enclave. Thus, we first determine an appropriate chunk size, for each of which the sealing function is executed.

The chunk size depends on time epochs, the enclave size, the computational overhead of executing sealing on the chunk, and the communication overhead for providing the chunk to the verifier. A small chunk size reduces the communication overhead and maintains the log minimality property, thereby during the log verification phase, a verifier retrieves only the desired log chunks, instead of retrieving the entire sensor data. Consequently, SP stores many chunks.

\medskip
\noindent\textbf{\textsc{Phase} 2: Hash-chain creation.} Consider a chunk, $\mathcal{C}_x$, that can store at most $n$ sensor readings, each of them of the format: $\langle d_i, s_j, t_k\rangle$. The sealing function checks each sensor reading against data-capture rules and adds sensor state to each reading, as: $\langle d_i, s_j, s_j.\mathit{state}, t_k\rangle$. Since in this section we assumed that all sensor data will be stored, the sensor state of each sensor reading is set to 1. The sealing function starts with the first sensor reading of the chunk $\mathcal{C}_x$, as follows:

\noindent\textit{\underline{First sensor reading}}. For the first sensor reading of the chunk, the sealing function computes a hash function on value zero, \textit{i}.\textit{e}, $H(0)$. Then, the sealing function mixes $H(0)$ with the remaining values of the sensor reading, \textit{i}.\textit{e}., sensor-id, device-id, sensor state, and time, at which it computes the hash function, denoted by $H(d_1||s_j|| s_j.\mathit{state}|| t_k||H(0))$ that results in a hash digest, denoted by $h_1^x$. After processing the complete first sensor reading of the chunk $\mathcal{C}_x$, the enclave writes cleartext first sensor reading of $\mathcal{C}_x$, \textit{i}.\textit{e}., $\langle d_1,s_j, s_j.\mathit{state}, t_k\rangle$ on the disk, which can be accessed by SP.

\noindent\textit{\underline{Second sensor reading}}. Let $\langle d_2, s_j, s_j.\mathit{state}, t_{k+1}\rangle$ be the second sensor reading. For this, the sealing function works identically to the processing of the first sensor reading. It computes a hash function on the second sensor values, while mixing it with the hash digest of the first sensor reading, \textit{i}.\textit{e}., $H(d_2||s_j|| s_j.\mathit{state}||t_{k+1}||h_1^x)$ that results in a hash digest, say $h_2^x$. Finally, the enclave writes the second sensor reading in cleartext on the disk.

\noindent\textit{\underline{Processing the remaining sensor readings}}. Likewise, for the second sensor reading processing, the sealing function computes the hash function on all the remaining sensor readings of the chunk $\mathcal{C}_x$. After processing the last sensor reading of the chunk $\mathcal{C}_x$, the hash digest $h_n^x$ is obtained.

\begin{figure*}[!t]
	\BB
	\begin{center}
		\includegraphics[scale=0.5]{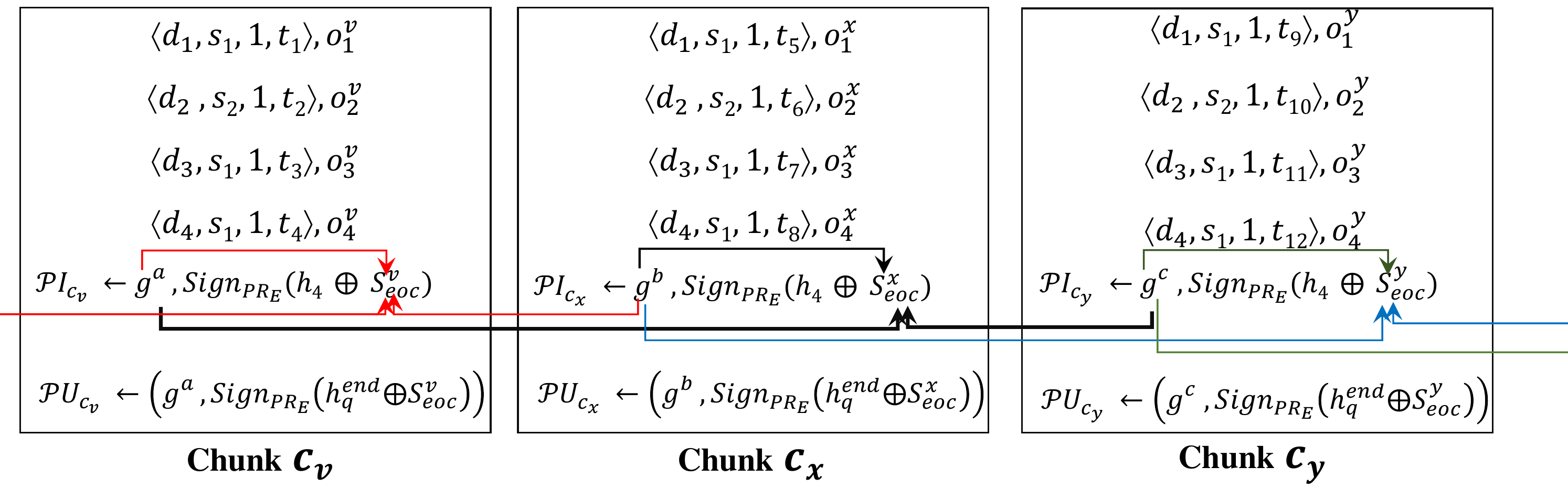}
	\end{center}
	\BB\B
	\caption{\textsc{Phase 3}: end of chunk, $S_{\mathit{eoc}}$, creation for three chunks. Observe that $S_{\mathit{eoc}}^{x} = g^a \oplus g^b \oplus g^c$.}
	\BBB
	\label{fig:eoc}
\end{figure*}

\medskip
\noindent\textbf{\textsc{Phase} 3: Proof-of-integrity creation.} Since each sensor reading is written on disk, SP can alter sensor readings, to make it impossible to verify log integrity by an auditor. Thus, to show that all the sensor readings are kept according to the pre-notified data-capture rules, the sealing function prepares an immutable proof-of-integrity for each chunk, as follows:

For each chunk $\mathcal{C}_i$, the sealing function generates a random string, denoted by $g^j$, where $i\neq j$. Let $\mathcal{C}_v$, $\mathcal{C}_x$, and $\mathcal{C}_y$ be three consecutive chunks (see Figure~\ref{fig:eoc}), based on consecutive sensor readings. Let $g^a$, $g^b$, and $g^c$ be random strings for chunks $\mathcal{C}_v$, $\mathcal{C}_x$, and $\mathcal{C}_y$, respectively. The use of random strings will ensure that any of the consecutive chunks have not been deleted by SP (will be clear in \S\ref{subsec:Attestation Phase}). Now, for producing the proof-of-integrity for the chunk $\mathcal{C}_x$, the sealing function: (\textit{i}) executes XOR operation on $g^a$, $g^b$, $g^c$, whose output is denoted by $S_{\mathit{eoc}}^{x}$, where $\mathit{eoc}$ denotes the end-of-chunk; (\textit{ii}) signs $h_n^x$ XORed with $S_{\mathit{eoc}}^{x}$ with the private key of the enclave; and (\textit{iii}) writes the proof-of-integrity for log verification of the chunk $\mathcal{C}_x$ with the random string $g^b$, as follows:
\centerline{
	$\mathcal{PI}_{\mathcal{C}_x}=(g^b,\mathit{Sign}_{\mathit{PR}_E}(h_n^x \oplus S_{\mathit{eoc}}^x))$
}

\smallskip\noindent\textbf{Note.} We do not generate the proof for each sensor reading. The enclave writes only the proof and the random string for each chunk to the disk, which is accessible by SP. Further, the sensor readings having the state one are written on the disk, based on which SP develops services.

\smallskip
\noindent\textbf{Example.} Please see Figure~\ref{fig:sealing function execution}, where the \textbf{middle box} shows \textsc{Phase} 2 execution on four sensor readings. Note that the hash digest of each reading is passed to the next sensor reading on which a hash function is computed with the sensor reading. After computing $h_4$, the proof-of-integrity, $\mathcal{PI}$, is created that includes signed $h_4\oplus S_{\mathit{eoc}}^x$ and a random string, $g^b$.

\smallskip
\noindent\textit{Note. $g^{\ast}$ for the first chunk.} The initialization of the log sealing function requires an initial seed value, say $g^{\ast}$, due to the absence of $0^{\mathit{th}}$ chunk. Thus, in order to initialize the secure binding for the first chunk, the seed value is used as a substitute random string.

\noindent\subsubsection{\textbf{\underline{Sealing Data for User Data/Service Verification}}}
\label{subsubsec:Sealing Data for Query Execution}
While capturing \emph{user-associated data}, users may wish to verify their user-associated data against notified messages. Note that \emph{the protocol presented so far requires entire cleartext data to be sent to the verifier to attest log integrity} (it will be clear soon in \S\ref{subsec:Attestation Phase}). However, such cleartext data transmission is not possible in the case of user-associated data verification, since it may reveal other users' privacy. Thus, to allow verification of user-associated data (or service/query result\footnote{{\scriptsize The users, who access services developed by SP (as mentioned in \S\ref{sec:introduction}), may also wish to verify the query results, since SP may tamper with the data to show the wrong results.}} verification), we develop a new sealing method, consists of the three phases: (\textit{i}) chunk creation, (\textit{ii}) hash-generation, and (\textit{iii}) proof-of-integrity creation. The chunk creation phase of this new sealing method is identical to the above-mentioned chunk creation phase 1; see \S\ref{subsubsec:Sealing Entire Sensor Data}. Below, we only describe \textsc{Phase 2} and \textsc{Phase} 3.


\medskip
\noindent\textbf{\textsc{Phase} 2: Hash-generation.} Consider a chunk, $\mathcal{C}_x$, that can have at most $n$ sensor readings, each of them of the format: $\langle d_i, s_j, s_j.\mathit{state}, t_k\rangle$. 
Our objective is to hide users' device-id and its frequency-count (\textit{i}.\textit{e}., which device-id is prominent in the given chunk). Thus, on the $i^{\mathit{th}}$ sensor reading, the sealing function mixes $d_j$ with $t_k$, and then, computes a hash function over them, denoted by $H(d_j||t_k)$ that results in a digest value, say $o_i$. Note that hash on device-ids mixed with time results in two different digests for more than one occurrence of the same device-id. Note that $o_i$ helps the user to know his presence/absence in the data during attestation, but it will not prove that tampering has not happened with the data. Then, the sealing function mixes $o_i$ with the sensor state (to produce a proof of sensor state) of the $i^{\mathit{th}}$ sensor reading, and on which it computes the hash function, denoted by $H(o_i||s_j.\mathit{state})$ that results in a hash digest, denoted by $\mathit{hu}_i^x$. After processing the $i^{\mathit{th}}$ sensor reading of the chunk $\mathcal{C}_x$, the enclave writes $o_i$ on the disk. After processing all the $n$ sensor readings of the chunk $\mathcal{C}_x$, the sealing function computes XOR operation on all hash digests, $\mathit{hu}_i^x$, where $1\leq i\leq n$: $\mathit{hu}_1^x\oplus \mathit{hu}_2^x \oplus \ldots \oplus \mathit{hu}_n^x$, whose output is denoted by $\mathit{hu}_{\mathit{end}}^x$. (Reason of computing $\mathit{hu}_{\mathit{end}}^x$ will be clear in \S\ref{subsec:Attestation Phase}).

\medskip
\noindent\textbf{\textsc{Phase} 3: Proof-of-integrity creation for the user.} The sealing function prepares an immutable proof-of-integrity for users, denoted by $\mathcal{PU}$, for each chunk and writes on the disk. Likewise, proof-of-integrity for entire log verification, $\mathcal{PI}$ (\S\ref{subsubsec:Sealing Entire Sensor Data}), for each chunk, the sealing function obtains $S_{\mathit{eoc}}$; refer to \textsc{Phase} 3 in \S\ref{subsubsec:Sealing Entire Sensor Data}. Now, for producing $\mathcal{PU}$ for the chunk $\mathcal{C}_x$, the sealing function: (\textit{i}) signs $\mathit{hu}^x_{\mathit{end}}$ XORed with $S_{\mathit{eoc}}^x$ with the private key of the enclave, and (\textit{ii}) writes the signed output with the random string of the chunk, $g^b$, as $\mathcal{PU}_{\mathcal{C}_x}$.

\centerline{
	$\mathcal{PU}_{\mathcal{C}_x}=(g^b,\mathit{Sign}_{\mathit{PR}_E}(\mathit{hu}^x_{\mathit{end}} \oplus S_{\mathit{eoc}}^x))$
}

\smallskip
\noindent\textbf{Note.} The enclave writes hash digests, $o_i$ for each sensor reading, the proof for user verification, and the random string for each chunk on the disk. Of course, the sensor readings having the state one are written on the disk.

\smallskip
\noindent\textbf{Example.} Please see Figure~\ref{fig:sealing function execution}, where the \textbf{last box} shows \textsc{Phase} 2 execution on four sensor readings to obtain the proof-of-integrity for the user, $\mathcal{PU}$. $\blacksquare$

{\color{black}

\subsubsection{\textbf{\underline{Sealing Mixed State Sensor Data}}}
\label{subsubsec:Sealing Partial-No Sensor Data}

The protocol so far has assumed that all data has the sensor state of one. We next consider how it can be generalized to a situation when some sensor readings may not satisfy the data capture rules (and hence, have the sensor state of zero). Recall that (as mentioned in \S\ref{sec:high_level_iot}), the enclave decrypts the sensor data received from the IFD and checks
against the pre-notified data-capture rules, and if a sensor reading captured by a sensor $s_i$ does not satisfy the data-capture rules, then the sensor state of the sensor reading becomes zero. Please note that the sensor state of zero does not indicate that the sensor is turned off.

\smallskip
\noindent
\textbf{Example~\ref{subsubsec:Sealing Partial-No Sensor Data}.1.} In a chunk, assume the following six sensor readings, produced by two sensors $s_1$ and $s_2$.

\centerline{$\langle d_1, s_1,1,t_1\rangle$}
\centerline{$\langle d_2, s_2,0,t_2\rangle$}
\centerline{$\langle d_2, s_2,0,t_3\rangle$}
\centerline{$\langle d_3, s_2,0,t_4\rangle$}
\centerline{$\langle d_3, s_2,1,t_5\rangle$}
\centerline{$\langle d_1, s_1,1,t_6\rangle$}

In this case, there is no need to store all sensor readings having a state zero. However, doing it carelessly may provide an opportunity to the adversary to delete all the sensor readings and prove that such readings are deleted due to the sensor state being zero, due to not satisfying data-capturing rules. Thus, to avoid storing sensor readings having sensor state zero, we provide a method below. $\blacksquare$

\medskip
Informally, the sealing function checks each sensor reading and deletes all those sensor readings for which sensor state is \texttt{passive}, due to not satisfying data-capture rules. In this case, it is not mandatory to seal each sensor readings, as mentioned in \S\ref{subsubsec:Sealing Entire Sensor Data}. Thus, the sealing function provides a {\em filter operation} that removes sensor readings whose device state is \texttt{passive}, while storing sensor readings whose device states are \texttt{active}. But, the sealing function cryptographically stores the data with minimal information of sensor readings whose device states are \texttt{passive}. Below we describe \textsc{Phase 2} for sealing mixed state sensor data. Note that \textsc{Phase 1} (chunk creation) and \textsc{Phase 3} (proof-of-integrity creation) are identical to the method described in \S\ref{subsubsec:Sealing Entire Sensor Data}.

\medskip
\noindent\textbf{\textsc{Phase} 2: Sealing operation.} Formally, to create hash-chain for this case, the sealing function will do the following: For the first $i^{\mathit{th}}$ sensor reading (for example, $\langle d_j,s_k,0,t_i\rangle$) whose $state=0$, the enclave executes two operations: (\textit{i}) sealing function that computes a hash function, as: $H(s_j||s_j.\mathit{state}|| t_k||h_{i-1})$, whose output is denoted by $h_i$, and (\textit{ii}) filter operation that deletes the device-id $d_j$ and stores $\langle s_k,0,t_i\rangle$ on the disk. Now for all the successive sensor readings until encountering a sensor reading, say $l$, with $state = 1$, all the sensor readings are discarded (not stored on the disk), as well as, the hash function is not executed. However, the sealing function computes the hash function on the $l^{\mathit{th}}$ sensor reading ($\langle d_x,s_k,1,t_l\rangle$), as: $H(d_x||s_j||1|| t_l||h_i)$ and stores $l^{\mathit{th}}$ sensor reading on the disk.

\smallskip\noindent
\textbf{Example~\ref{subsubsec:Sealing Partial-No Sensor Data}.2.} Now, we apply the above-mentioned \textsc{Phase} 2 on the sensor readings given in Example~\ref{subsubsec:Sealing Partial-No Sensor Data}.1. Here, the enclave does not store the second, third, and fourth sensor readings on disk and delete them, while the remaining sensor readings will be stored on the disk. In addition, the enclave will store only the sensor device and its state of the first reading (\textit{i}.\textit{e}., the second sensor reading) for which state was zero and generate verifiable logs, such that the verifier can verify that the three entries has been deleted due to not satisfying  data-capturing rules. The sealing function computes the hash-chain and proof-of-integrity as follows:

\centerline{$ h_1\leftarrow H(d_1||s_1||1||t_1||H(0))$}
\centerline{$h_2\leftarrow H(s_2      ||0||t_2||h_1)$}
\centerline{$h_3\leftarrow H(d_3 ||s_2||1||t_5||h_2)$}
\centerline{$h_4\leftarrow H(d_1 ||s_1||1||t_6||h_3)$}
\centerline{
	$\mathcal{PI} \leftarrow \langle \mathit{secret\: string}, \mathit{sign}_{\mathit{PR}_E}(\mathit{secret\: string} \oplus h_4)\rangle$}
Note that here we compute a hash function on the first, second, fifth, and last sensor readings, while do not seal the third and fourth sensor readings, due to their passive (or zero) sensor states.\footnote{{\scriptsize We can only compress $x>1$ continuous sensor readings, say $\langle \ast_d, \ast_s, \ast_s.\mathit{state}=0,\ast_t\rangle$ (where $\ast_d$, $\ast_s$, and $\ast_t$ denote any device-id, sensor-id, and time, respectively) to produce a proof that such $x$ readings have been deleted. However, we cannot compress $x$ sensor readings having $\ast_s.\mathit{state}=1$, since it disallows to verify service integrity (\textit{e}.\textit{g}., a user query, how many times a user has visited a location, cannot be verified, if $x$ readings with $\ast_s.\mathit{state}=1$ have been deleted).}} $\blacksquare$ 

\noindent
\subsubsection{\textbf{\underline{Log-size Optimization}}}
\label{subsubsec:optimization}
The above-mentioned procedure given in \S\ref{subsubsec:Sealing Partial-No Sensor Data} regards the sensor states and stores less amount of cryptographically sealed data on the disk, by not storing consecutive sensor readings having sensor state of zero. However, such improvement is limited, if the sensor readings have state zero and one in an alternative sequence.

\smallskip
\noindent
\textbf{Example~\ref{subsubsec:optimization}.1.} Consider the following seven sensor readings obtained from two sensors $s_1$ and $s_2$. Here, the state of sensor $s_2$ is zero, due to not satisfying the data-capturing rules; however, the state of sensor $s_1$ is one.

\centerline{$\langle d_1, s_1,1,t_1\rangle$}
\centerline{$\langle d_2, s_2,0,t_2\rangle$}
\centerline{$\langle d_2, s_1,1,t_3\rangle$}
\centerline{$\langle d_3, s_2,0,t_4\rangle$}
\centerline{$\langle d_3, s_1,1,t_5\rangle$}
\centerline{$\langle d_1, s_2,0,t_6\rangle$}
\centerline{$\langle d_1, s_1,1,t_7\rangle$}

In such scenario, the method given in \S\ref{subsubsec:Sealing Partial-No Sensor Data} is not efficient, since it will store all the sensor readings and produce hash digest for each sensor reading, as follows:

\centerline{$h_1\leftarrow H(d_1||s_1||1||t_1||H(0))$}
\centerline{$h_2\leftarrow H(s_2      ||0||t_2||h_1)$}
\centerline{$h_3\leftarrow H(d_2 ||s_1||1||t_3||h_2)$}
\centerline{$h_4\leftarrow H(s_2      ||0||t_4||h_3)$}
\centerline{$h_5\leftarrow H(d_3 ||s_1||1||t_5||h_4)$}
\centerline{$h_6\leftarrow H(s_2      ||0||t_6||h_5)$}
\centerline{$h_7\leftarrow H(d_1 ||s_1||1||t_7||h_6)$}

Thus, in order to reduce the size of the secured log for the case when sensor readings have state zero and one in an alternative sequence, below, we propose a method that stores logs for each sensor or each user device. $\blacksquare$

\medskip
\noindent
\textbf{Per sensor-based logging.} We implement a log-sealing procedure (given in \S\ref{subsubsec:Sealing Entire Sensor Data} and \S\ref{subsubsec:Sealing Partial-No Sensor Data}) on per sensor, \textit{i}.\textit{e}., we group the sensor readings produced by the same sensor before producing cryptographically sealed logs. This optimization method works as follows, for $n$ sensor readings and $x$ sensors:
\begin{enumerate}[noitemsep,nolistsep,leftmargin=0.2in]
	\item The enclave creates $x$ buffers, one buffer for each of the $x$ sensors.
	\item The enclave reads the $n$ sensor readings from the disk and decrypts them.
	\item Each of the $n$ sensor readings is allocated to one of the buffers based on the sensor.
	\item On each sensor reading allocated to a buffer, the enclave either executes the method given in \S\ref{subsubsec:Sealing Entire Sensor Data} if all the sensor readings have state one, or executes the method given in \S\ref{subsubsec:Sealing Partial-No Sensor Data} if the sensor readings have state one and zero.
	\item At the end, the enclave writes cryptographically secured logs and proof-of-integrity for each buffer's sensor log, on the disk.
\end{enumerate}

\smallskip
\noindent
\textbf{Example~\ref{subsubsec:optimization}.2.} Now, we show how one can reduce the size of secured logs for the seven sensor readings given in Example~\ref{subsubsec:optimization}.1 by keeping logs per sensor-based.
Here, the enclave maintains two buffers: one for $s_1$ and another for $s_2$. All the sensor readings of $s_1$ are sealed using the method of \S\ref{subsubsec:Sealing Entire Sensor Data}, and the sensor readings of $s_2$ are sealed using the method of \S\ref{subsubsec:Sealing Partial-No Sensor Data}. Thus, the enclave executes the following computation:

\noindent The first buffer dealing with the sensor readings of sensor $s_1$:\\
\centerline{$h_1\leftarrow H(d_1 ||s_1||1||t_1||H(0))$}
\centerline{$h_2\leftarrow H(d_2 ||s_1||1||t_3||h_1)$}
\centerline{$h_3\leftarrow H(d_3 ||s_1||1||t_5||h_2)$}
\centerline{$h_4\leftarrow H(d_1 ||s_1||1||t_7||h_3)$}
\centerline{
	$\mathcal{PI}_{s_1} \leftarrow \langle \mathit{secret\: string}, \mathit{sign}_{\mathit{PR}_E}(\mathit{secret\: string} \oplus h_4)\rangle$}

\noindent The second buffer dealing with the sensor readings of sensor $s_2$:\\
\centerline{$h_{5}\leftarrow H(s_2      ||0||t_2||H(0))$}
\centerline{
	$\mathcal{PI}_{s_2} \leftarrow \langle \mathit{secret\: string}, \mathit{sign}_{\mathit{PR}_E}(\mathit{secret\: string} \oplus h_5)\rangle$}

Note that, here, we used the notations $\mathcal{PI}_{s_1}$ and $\mathcal{PI}_{s_2}$ to indicate the proof-of-integrity generated for sensors $s_1$ and $s_2$, respectively. The enclave writes cleartext sensor readings corresponding to the sensor $s_1$, the tuple $\langle s_2, 0, t_2\rangle$, hash digests $h_1, h_2, \ldots h_5$, and proof-of-integrity for both sensors $\mathcal{PI}_{s_1}$ and $\mathcal{PI}_{s_2}$, on the disk. $\blacksquare$

\smallskip
\noindent
\textbf{Issues.} The above-mentioned method, while reduces the size of cryptographically sealed logs, it faces two issues, as follows:
\begin{enumerate}

	\item \emph{More sensors.} It may happen that the number of sensors is significantly more, thereby it is not easy to maintain buffers for each sensor, due to a limited memory of the secure hardware. In this case, we can still use the above method; however, one buffer is responsible for more than one sensor. For example, if there are four sensors, $s_1,s_2,\ldots,s_4$, and the secure hardware can hold only two buffers, then the first buffer might be responsible for sensor readings corresponding to two sensors $s_1$ and $s_2$, and another buffer might be responsible for sensor readings corresponding to the remaining two sensors. To allocate sensor readings to buffers, one can use a hash function on the sensor-id to know the buffer id.

	\item \emph{Different-sized log.} Note that when we create multiple buffers for sensors, the enclave writes multiple chunks and proof-of-integrity corresponding to multiple buffers. The size of each chunk may not be identical, due to having a different number of hash digests. It may reveal information to the adversary that which sensor's state is zero. To avoid such leakage, the enclave may pad output produced for each buffer with some fake values and may include this information in proof-of-integrity to show how many fake values are added. For instance, in Example~\ref{subsubsec:optimization}.2, the output corresponding to the second buffer has only one hash digest (\textit{i}.\textit{e}., $h_5$) while the output corresponding to the first buffer will have four hash digests (\textit{i}.\textit{e}., $h_1,h_2,h_3,h_4$). Thus, to write the same size data for each buffer, the enclave may pad the output corresponding to the second buffer with three fake hash digests and write this information in the proof-of-integrity, as follows: $\mathcal{PI}_{s_2} \leftarrow \langle \mathit{secret\: string}, \mathit{sign}_{\mathit{PR}_E}(\mathit{secret\: string} \oplus h_5, 3)\rangle$. Note that while adding fake hash digests, the user does not need to verify any fake digest, and thus, verifying only desired data will reduce the verification time, as we will see in Experiment 7 in \S\ref{sec:Experimental Evaluation}.
\end{enumerate}

\noindent
\textbf{Note: Per user-based logging.} We can also apply the same procedure for each user device-id to reduce the size of the cryptographically sealed data, by creating buffers for each user device id. This may reduce the verification time for user-related data.

}

\subsection{Attestation Phase}
\label{subsec:Attestation Phase}
The attestation phase contains two sub-phases: (\textit{i}) key establishment between the verifier and service provider to retrieve logs (\S\ref{subsubsec:Key Establishment}), and (\textit{ii}) verification of the retrieved logs (\S\ref{subsubsec:Verification of Logs}).

\begin{figure}[!t]
	\begin{center}
		\includegraphics[scale=0.3]{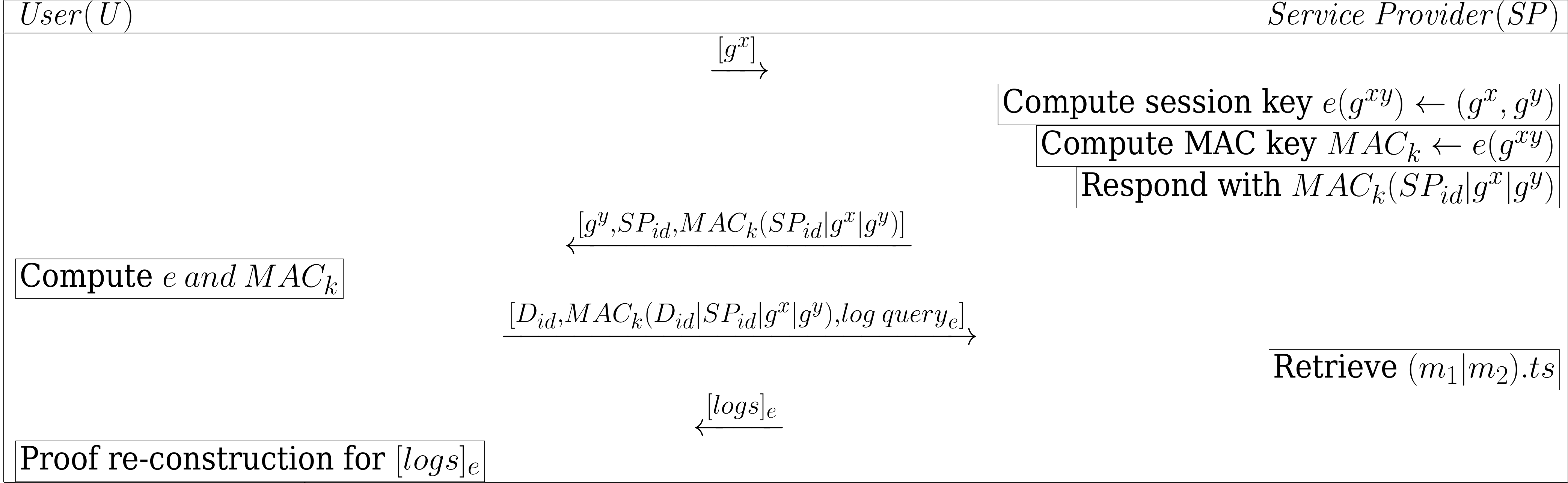}
	\end{center}
	\vspace{-0.2cm}
	\caption{{\color{black} Secure log retrieval by the verifier.}}
	\label{fig:esro}
	\BBB
\end{figure}

{\color{black}
\subsubsection{\textbf{\underline{Key Establishment}}}
\label{subsubsec:Key Establishment}
The secure log retrieval is crucial for proof validation and non-trivial in the presence of a de-centralized verifier model, where anyone can execute a remote request to attest the secured logs against the data-capture rules.
Thus, before the log transmission from the service provider to the verifier (\textit{i}.\textit{e}., a user or an auditor), for each attestation request, the verifier's identity must be authenticated prior to the session key establishment.

Our log retrieval scheme is based on Authenticated Key Exchange (AKE) protocol~\cite{luks}, where a verifier and the service provider (\textit{i}.\textit{e}., prover) dynamically establish a session key (shown in Figure~\ref{fig:esro}). This dynamic key establishment provides forward secrecy for all future sessions, such that, for any compromised session in the future, all sessions in the past remain secure. To achieve these properties, we use SIGMA~\cite{luks} protocol, which is the cryptographic base for the Internet Key Exchange (IKE) protocol. The family of SIGMA protocols is based on Diffie-Hellman key exchange~\cite{DBLP:journals/tit/DiffieH76}. We only show a 3-round version of SIGMA, as it provides verifier/sender-identity protection during the key establishment process. Recall that in our solution, the prover is a centralized service provider, but the verifier can be anyone, and therefore, we use this identity protection method to achieve the verifier's identity privacy during the session.

Without the loss of generality, we, first, define the Computational Diffie Hellman (CDH)~\cite{DBLP:journals/tit/DiffieH76}. During the session, the verifier and the prover must execute the computations within a cyclic group $G=\langle g \rangle$ that has a generator $g$ of prime order $q$. According to the CDH assumption, the computation of a discrete logarithm function on public values $(g,g^x,g^y)$ is hard within the cyclic group $G$. In particular, given the publicly known value $g$, it is hard to distinguish $g^{xy}$ from $g^x$ and $g^y$ without knowing the ephemeral secrets $x$ and $y$.

Figure~\ref{fig:esro} depicts SIGMA-based communication flow between a verifier and the service provider/prover. Initially, a verifier selects a secret value $x$, computes $g^x$ as a public exponent, and sends it to the prover. Similarly, the prover selects a secret value $y$, computes $g^y$, and receives the public exponent $g^x$ from the verifier. Next, the prover computes a joint secret value as: $e=g^{xy}$, and also uses it to derive a message authentication code (MAC) key $\mathit{MAC}_k$.

The prover composes a message structure as: $[g^y,\mathit{SP}_{id},\mathit{MAC}_k(\mathit{SP}_{id}||g^x||g^y)]$, where $g^y$ is the prover's public exponent for session key generation, $\mathit{SP}_{id}$ is the identity of sender/prover, and $\mathit{MAC}_k(\mathit{SP}_{id}||g^x||g^y)$ is the message authentication code generated on sender's identity and all public exponents used for session key derivation. Note that in order to generate this MAC, a separate key $\mathit{MAC}_k$ is derived from the session key $e$.

Next, the verifier receives this message and retrieves the public exponent $g^y$ to generate a local copy of session key $e$, as well as, the message authentication code generating key $\mathit{MAC}_k$ derived from $e$. At this stage, both parties have locally computed a secure session key $e$; however, there is still an identity disclosure required from the verifier to prove that the freshly generated key $e$ binds to an authentic identity holder $v_{id}$, where $v_{id}$ is the verifier's identity.

Thus, the verifier sends a message, consisting of $[v_{id},\mathit{MAC}_k(v_{id}||\mathit{SP}_{id}||g^x||g^y),\langle\mathit{log} \: \mathit{query}\rangle_e]$, where $\mathit{MAC}_k(v_{id}||\mathit{SP}_{id}||g^x||g^y)$ is the message authentication code on verifier's identity $v_{id}$, responder's identity $\mathit{SP}_{id}$, and all public exponents exchanged so far, and $\langle\mathit{log} \: \mathit{query}\rangle_e$ is the log query request protected by the freshly generated session key $e$. This step binds all public key exponents exchanged with the claimed identity holders together and marks the end of the authenticated session key exchange process. The service provider, then, retrieves the corresponding logs according to the log query request in the last message and sends the log encrypted using the session key $e$ to the verifier.

}

\subsubsection{\textbf{\underline{Verification of Logs}}}
\label{subsubsec:Verification of Logs}
This section presents procedures for log verification at the auditor and a user.

\noindent\textbf{Verification process at the auditor.} Recall that the auditor can verify any part of the sensor data. Suppose the auditor wishes to verify a chunk $\mathcal{C}_x$; see Figure~\ref{fig:eoc}. Hence, entire sensor data (the data written in the first box of Figure~\ref{fig:sealing function execution}) of the chunk $\mathcal{C}_x$, random strings $g^a$, $g^b$, and $g^c$ (corresponding to the previous and next chunks of $\mathcal{C}_x$; see Figure~\ref{fig:eoc}), and proof-of-integrity $\mathcal{PI}_{\mathcal{C}_x}$ are provided to the auditor. The auditor performs the same operation as in \textsc{Phase} 2 of \S\ref{subsubsec:Sealing Data for Query Execution}. Also, the auditor computes the end-of-chunk string $S^x_{\mathit{eoc}}=g^a\oplus g^b\oplus g^c$. In the end, the auditor matches the results of $h^x_n \oplus S^x_{\mathit{eoc}}$ against the decrypted value of received $\mathcal{PI}_{\mathcal{C}_x}$, and if both the values are identical, then it shows that the entire chunk is unchanged.

Note that since SP transfers sensor readings of the chunk $\mathcal{C}_x$, random strings ($g^a$, $g^b$, and $g^c$) and $\mathcal{PI}_{\mathcal{C}_x}$ to the user, SP can alter any transmitted data. However, SP cannot alter the signed $\mathit{Sign}_{\mathit{PR}_E}(h_n^x \oplus S_{\mathit{eoc}}^x)$, due to the unavailability of the private key of the enclave, $\mathit{PR}_E$, which was generated and provided by the trusted authority to the enclave. Thus, by following the above-mentioned procedure on the sensor readings of $\mathcal{C}_x$, any inconsistency created by SP will be detected by the auditor.

\smallskip
\noindent\textbf{Verification process at the user.} If the user wishes to verify his data in a chunk, say $\mathcal{C}_x$, the user is provided all hash digests computed over device-id and time ($o_i$, see the last box in Figure~\ref{fig:sealing function execution}), time, sensor state, random strings $g^a$, $g^b$, and $g^c$ (see Figure~\ref{fig:eoc}), and the proof $\mathcal{PU}$ by SP. Since, the user knows her device-id, first, the user verifies her occurrences in the data by computing the hash function on her device-id mixed with provided time values and compares against received hash digests. This confirms the user's presence/absence in the data. Also, to verify that no hash-digest is modified/deleted by SP, the user computes the hash function on the sensor state mixed with the received $o_i$ ($1\leq i\leq n$, where $n$ in the number of sensor readings in $\mathcal{C}_x$) and computes $\mathit{hu}_{\mathit{end}}^x = h_1^x\oplus h_2^x \oplus \ldots \oplus h_n^x$. Finally, the user computes $\mathit{hu}_{\mathit{end}}^x \oplus S^x_{\mathit{eoc}}$ and compares against the decrypted value of $\mathcal{PU}$. The correctness of this method can be argued in a similar manner to the correctness of the verification at the auditor.

{\color{black}
\section{System Implementation and Deployment}
\textsc{IoT Notary} has been implemented and deployed as an integral part of TIPPERS that is an IoT testbed at the University of California Irvine, providing various real-time location-based services on the campus. The university IT department is a trusted infrastructure deployer in this testbed. At UCI, users need to register their devices' MAC addresses in a system shown in Figure~\ref{fig:wifi-registration} to get access to the campus-wide WiFi network service. When a user's device connects to the campus WiFi network infrastructure, an association event, including the user device MAC address, access point ID, and timestamp of the connection, is generated by the access point and collected by the IT department as part of the infrastructure maintenance tasks. The \textsc{IoT Notary} protocol is used when a service provider, \textit{e}.\textit{g}., a location-based service in TIPPERS, asks for the WiFi association data for service provisioning.

Before sharing data with a service provider, \textit{i}.\textit{e}., TIPPERS, the university IT department sends the notifications to the public through emails, website messages (Figure~\ref{fig:website-notification}), and signage (Figure~\ref{fig:signage}) attached to the buildings where the data is collected. The IT department generates a secret key and corresponding SGX enclave program at the setup phase for each service provider. The SGX program is given to the service provider and runs on its SGX-enabled system. In operation, the IT department encrypts the WiFi association data using the secret key and sends the encrypted data to the service provider's SGX enclave program. The IT department is also responsible for policy management in our testbed as the university authority requires. A consent management website shown in Figure~\ref{fig:opt-out} is provided to users on campus to opt-out or opt-in their devices for data sharing with TIPPERS.

\begin{figure*}
        \centering
        \begin{subfigure}[b]{0.475\textwidth}
            \centering
            \includegraphics[width=\textwidth]{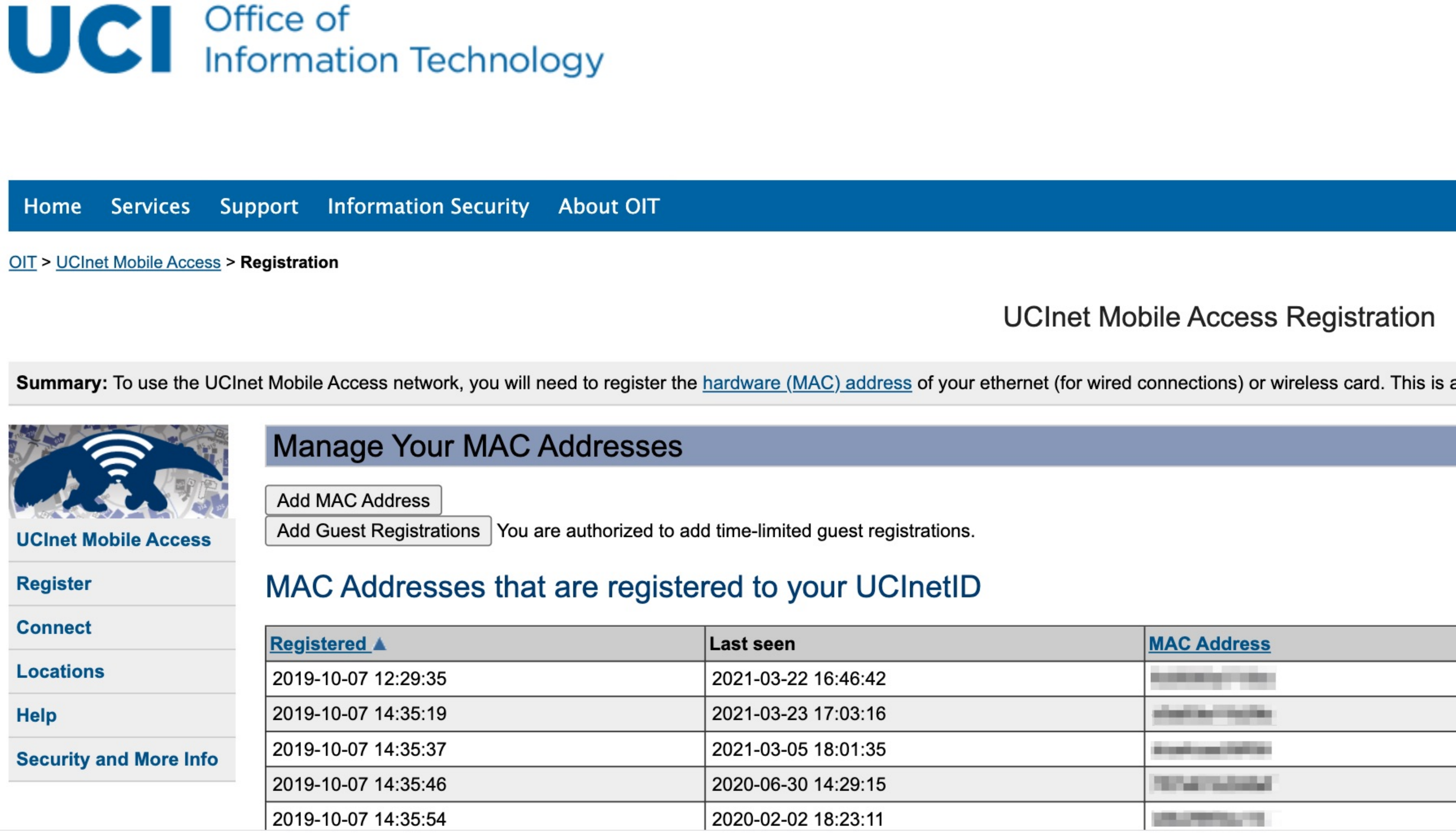}
            \caption[]%
            {{\color{black}{\small System for registering device's MAC address to get WiFi access on campus}}}
            \label{fig:wifi-registration}
        \end{subfigure}
        \hfill
        \begin{subfigure}[b]{0.475\textwidth}
            \centering
            \includegraphics[width=\textwidth]{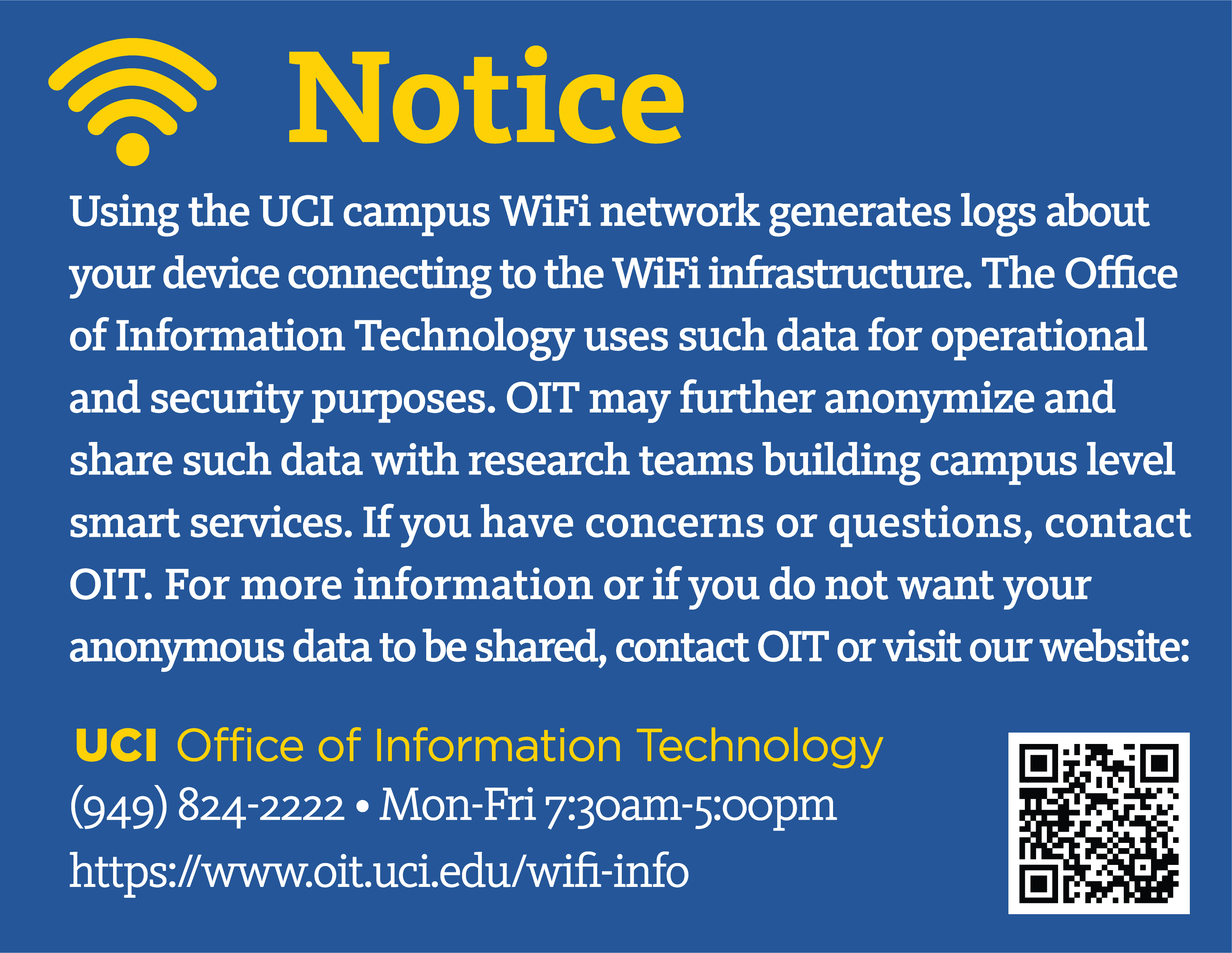}
            \caption[]%
            {{\color{black}{\small The data-capturing notification signage}}}
            \label{fig:signage}
        \end{subfigure}
        \vskip\baselineskip
        \begin{subfigure}[b]{0.475\textwidth}
            \centering
            \includegraphics[width=\textwidth]{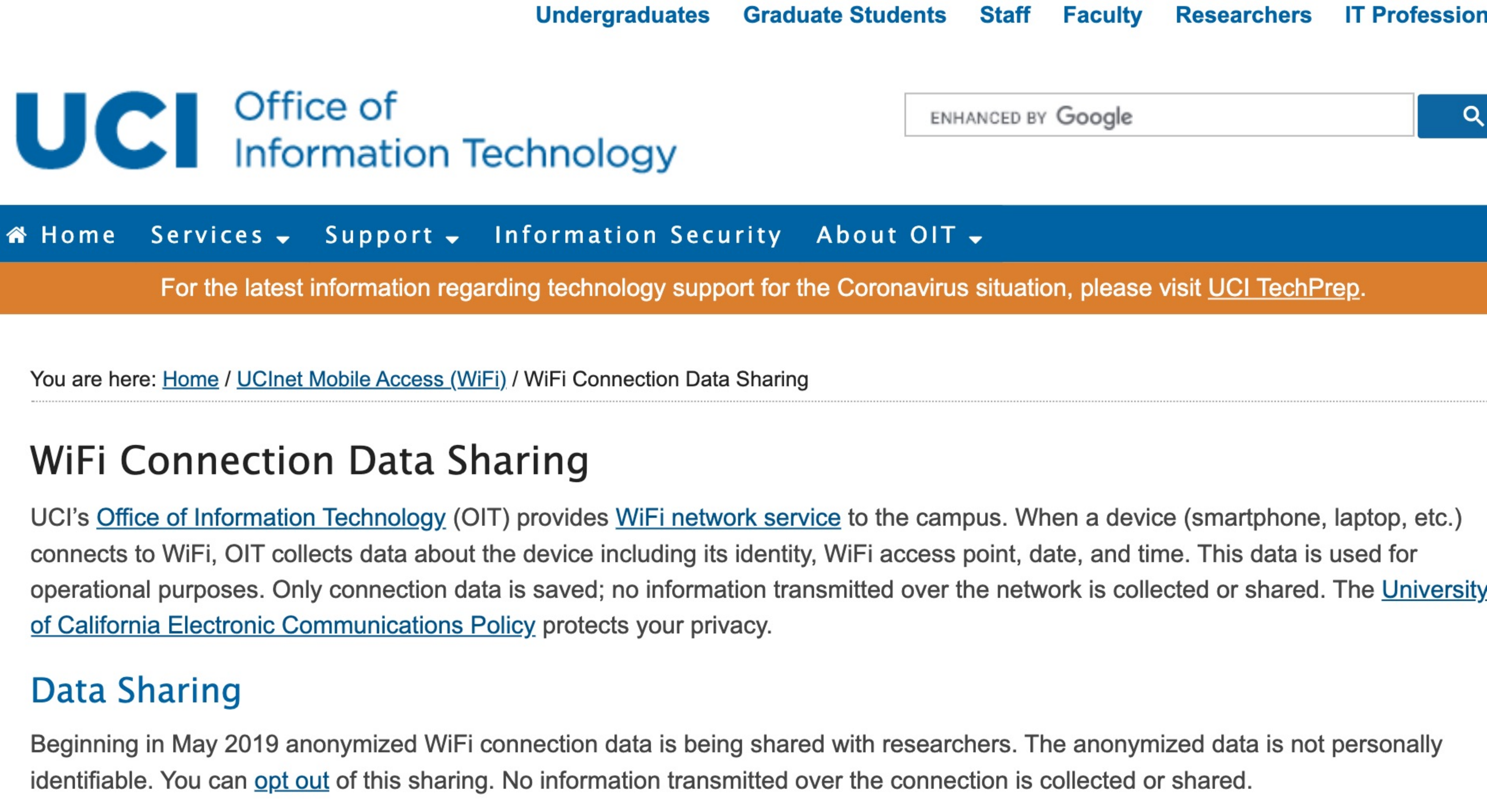}
            \caption[]%
            {{\color{black}{\small The data-capturing notification on IT department website}}}
            \label{fig:website-notification}
        \end{subfigure}
        \hfill
        \begin{subfigure}[b]{0.475\textwidth}
            \centering
            \includegraphics[width=\textwidth]{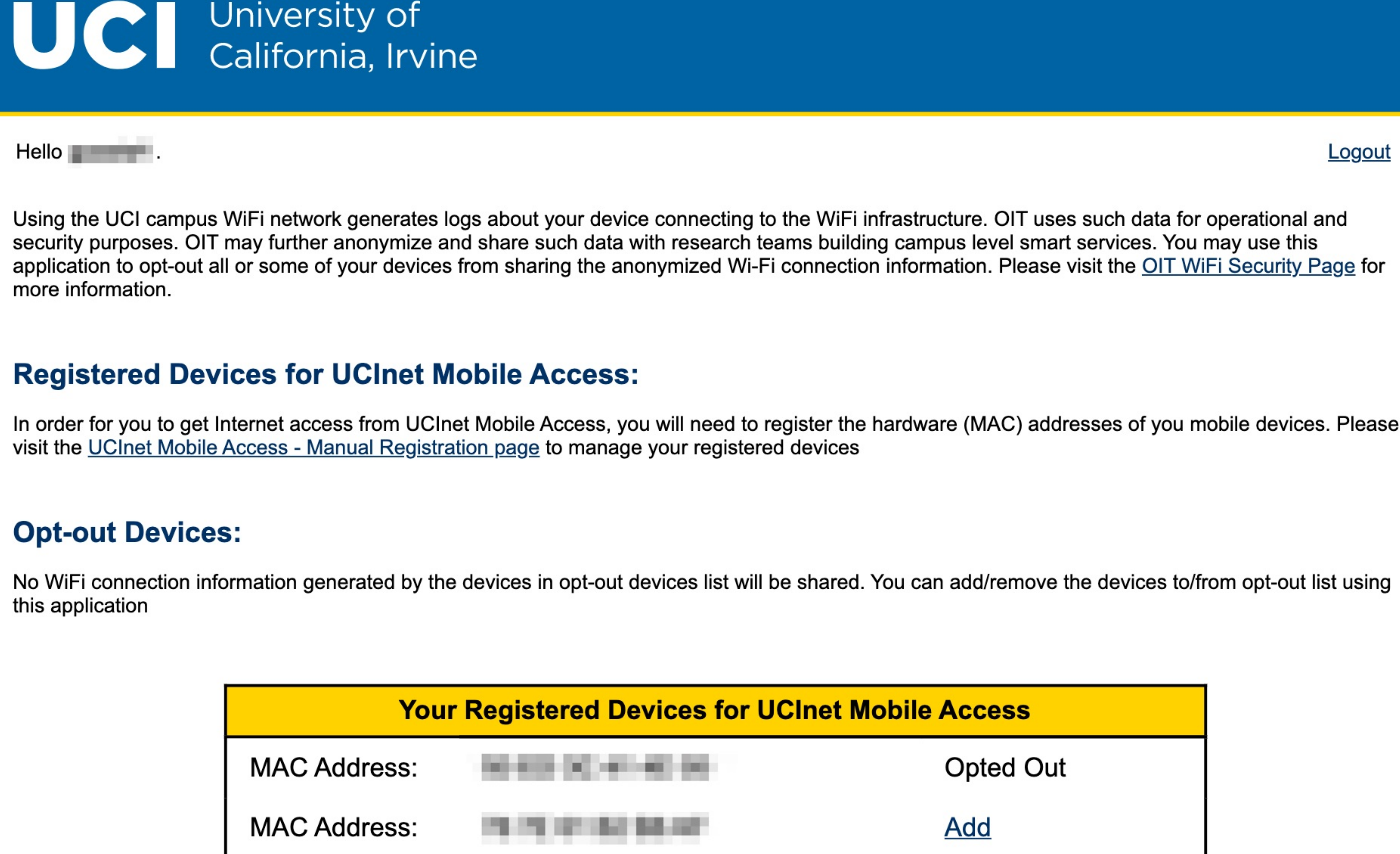}
            \caption[]%
            {{\color{black}{\small System for users to specify their policies (opt-out)}}}
            \label{fig:mean and std of net44}
        \end{subfigure}
        \caption[  ]
        {\color{black}{\small The campus deployment of \textsc{IoT Notary}}}
        \label{fig:opt-out}
    \end{figure*}

When the encrypted data arrives at TIPPERS, the SGX program decrypts the data in the secure enclave. It, further, enforces the policies by removing the data that contains opted-out devices. Then, tamper-proof logs are generated in the process and stored in a database. Before giving the data to TIPPERS' applications, the SGX program further de-identifies the MAC address.
Then, TIPPERS uses the data and provides valuable location-based services, such as occupancy analysis.
\textsc{IoT Notary} via TIPPERS provides APIs for the auditor and the users to check the presence/absence of their data by retrieving the desired logs for a given duration and an application for executing the attestation module at their-ends.

}

\section{Experimental Evaluation}
\label{sec:Experimental Evaluation}
This section presents our experimental results on live WiFi data. We execute \textsc{IoT Notary} on a 4-core 16GB RAM machine equipped with SGX at Microsoft Azure cloud.

\medskip
\noindent\textbf{Setup.} In our setup, the IT department at UCI is the trusted infrastructure deployer. It also plays the role of the trusted notifier (notifying users over emailing lists). At UCI, 490 WiFi sensors, installed over 30 buildings, send data to a controller that forwards data to the cloud server, where \textsc{IoT Notary} is installed. The cloud keeps cryptographic log digests that are transmitted to the verifier, while sensor data, qualifies data-capture rules, is ingested into real-time applications supported by TIPPERS. We use SHA-256 as the hashing algorithm and 256-bit length random strings in \textsc{IoT Notary}. We allow users to verify the data collected over the last 30minutes (min). 

\medskip\noindent\textbf{Dataset size.} Although \textsc{IoT Notary} deals with live WiFi data, we report results for data processed by the system over 180 days during which time \textsc{IoT Notary} processed 13GB of WiFi data having 110 million WiFi events.

\medskip\noindent\textbf{Data-capture rules.} We set the following four data-capture rules: (\textit{i}) \textit{Time-based}: always retain data, except from $t_i$ to $t_j$ each day; (\textit{ii}) \textit{User-location-based}: do not store data about specified devices if they are in a specific building; (\textit{iii}) \textit{User-time-based}: do not capture data having a specific device-id from $t_x$ to $t_y$ ($x\neq i$, $y\neq j$) each day; and (\textit{iv}) \emph{Time-location-based}: do not store any data from a specific building from time $t_x$ to $t_y$ each day. The validity of these rules was 40 days. After each 40-days, variables $i$, $j$, $x$, $y$ were changed.

\begin{figure}[!t]
	\centering
	\includegraphics[scale=0.5]{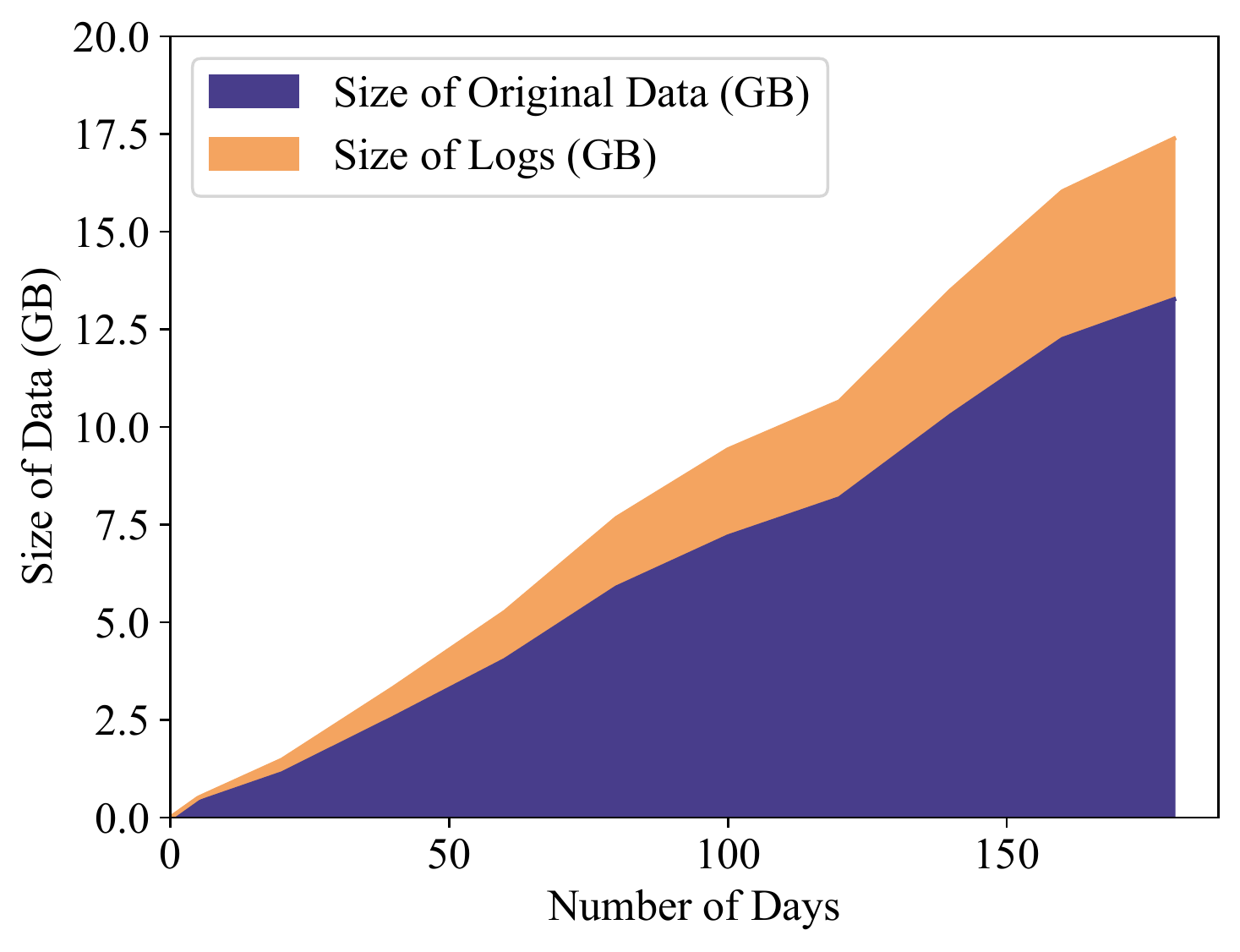}
	\caption{\color{black}{Exp 1: Storage overhead.}}
	\label{fig:Storage overhead}
	\BBB
\end{figure}

\medskip\noindent\textbf{Exp 1. Storage overhead at the cloud.} We fix the size of each chunk to 5MB, and on average, each of them contains $\approx$ 37K sensor readings, covering around 30min data of 30 buildings in peak hours. Based on the 5MB chunk size, we got 3291 chunks for 180 days. For each chunk, the sealing function generates two types of logs: (\textit{i}) for auditor verification that produced proof-of-integrity $\mathcal{PI}$ of size 512bytes, and (\textit{ii}) for user verification that produces hashed values (see Figure~\ref{fig:sealing function execution}) and proof-of-integrity for users $\mathcal{PU}$ of size 1.05MB. Figure~\ref{fig:Storage overhead} shows 180-days WiFi data size without having sealed logs ({\color{black}dark blue color}) and with sealed logs ({\color{black}light brown color}).\footnote{{\scriptsize The reason for getting more chunks is that during non-peak hours a 5MB chunk can store sensor readings for more than one hour. However, as per our assumption, we allow the user to verify the data collected over the last 30min. Hence, regardless of the chunk is full or not, we compute the sealing function on each chunk after 30min.}}

\medskip\noindent\textbf{Exp 2. Performance at the cloud.} For each 5MB chunk, the sealing function took around 310ms to seal each chunk. This includes time to compute $\mathcal{PI}$, $\mathcal{PU}$ and encrypt them. 

\smallskip
\noindent\textbf{Exp 3. Auditor verification time.} The auditor at our campus has a 7th-Gen quad-core i7CPU and 16GB RAM machine. It downloads the chunks from the cloud and executes auditor verification. The auditor varied the number of chunks from 1 to 3000; see Table~\ref{tbl:auditor_verification_time_cloud}. Note that to attest one-day data across 30 buildings, the auditor needs to download at most 50 chunks, which took less than 1min to verify. Observe that as the number of chunks increases, the time also increases, due to executing the hash function on more data.

\begin{table}[!h]
	\centering
	\begin{tabular}{|l|l|l|l|l|l|l|l|}
		\hline
		Number of Chunks  & 1   &50 &  100 & 500 & 1000 &  3000\\ \hline
		$\approx$ duration (day) & 30-60min & 1-2 & 2-5 & 8-18 & 35-55 & 175\\ \hline
		Verification time (seconds) & 1 & 49& 102 & 544 & 1160 & 4400 \\ \hline
	\end{tabular}
	\caption{Exp 3: The auditor verification time. Duration varies due to different class schedules in buildings and working hours.}
	\label{tbl:auditor_verification_time_cloud}
\end{table}

\begin{figure}[!t]
	\centering
	\includegraphics[scale=0.6]{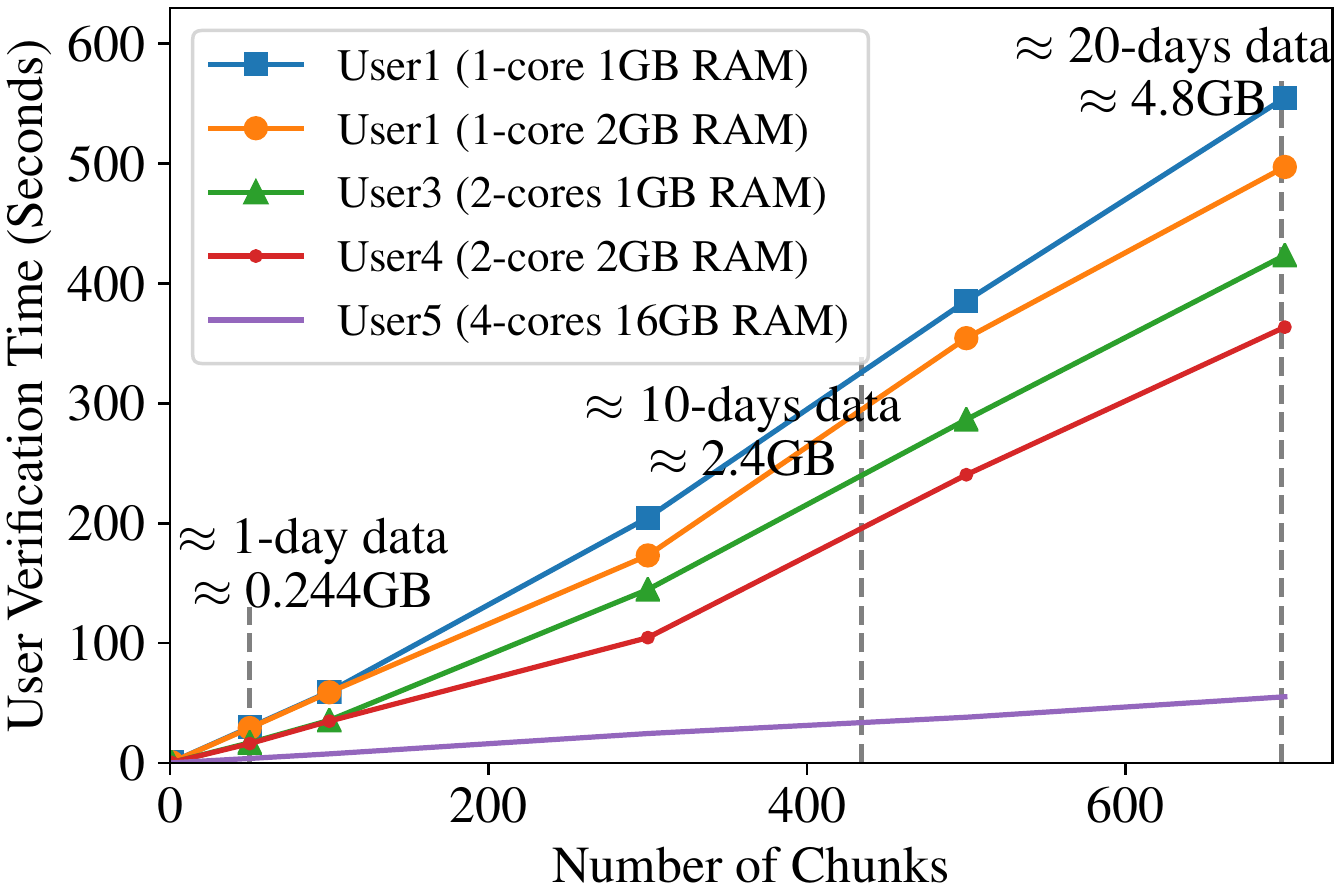}
	\caption{Exp 4: Verification time.}
	\label{fig:User verification time}
\end{figure}

\medskip\noindent\textbf{Exp 4: Verification at a resource-constrained user.} To show the practicality of \textsc{IoT Notary} for resource-constrained users, we considered four types of users, differing on computational capabilities (\textit{e}.\textit{g}., available main memory (1GB/2GB) and the number of cores (1 or 2 cores)). Each user verified 1/10/20-days data; see Figure~\ref{fig:User verification time}. Note that verifying 1-day data, which is $\approx$ 50 blocks, at resource-constrained users took at most 30s. As the number of blocks increases, the computational time also increases, where the maximum computational time to verify 20-days data was $<$ 10min. As the days increase, so does data transmitted to the user, which spills over to disk causing an increased latency. Also, we execute the same experiment on a powerful user having 4-core and 16GB machine. Note that as the number of core and memory increases, it results in parallel processing and the absence of disk data read. Thus, the computation time decreases (see user 5 in Figure~\ref{fig:User verification time}).

\begin{table}[!h]
{\color{black}
	\centering
	\begin{tabular}{|l|l|l|l|l|l|l|l|l|}\hline
		Chunk Size (KB)    & 5  & 100 & 1,000 & 5,000 & 10,000 & 25,000 & 40,000 \\\hline
		Time (millisecond) & 30 & 38  & 120   & 246   & 579    & 1,789  & 3,389 \\\hline
	\end{tabular}
	\caption{{\color{black} Exp 5: Impact of the chunk size on log sealing execution.}}
	\label{tab:Impact of chunk size on log sealing execution}
}
\end{table}

{\color{black}

\medskip\noindent\textbf{Exp 5. Impact of the chunk size.} We have selected chunk size to be 5MB that can hold at most 30min WiFi data. Now, we investigate the impact of chunk size on the sealing function execution time, the verification time, and latency in obtaining the most recent data.
Table~\ref{tab:Impact of chunk size on log sealing execution} shows that as the chunk size increases, the chunk holds more data, and hence, executing the sealing function on large-sized data takes more time.
Similarly, a large-sized chunk also increases the latency in obtaining the most recent data (see Figure~\ref{fig:latency vs log size}), since unless filling the chunk, the enclave cannot produce the proof-of-integrity. 
We can also create a chunk having only a single row; however, it will increase the size of secured logs (see Figure~\ref{fig:latency vs log size}). Observe that (in Figure~\ref{fig:latency vs log size}), when the chunk size is 5KB, the secured log size is $\approx$ 7GB, while when the chunk size is 5MB, the secured log size is $\approx$ 4GB. Figure~\ref{fig:plot-perud-vs-logisze} shows verification time also increases as the chunk size increases.

\begin{figure}[!t]
	\centering
	\includegraphics[scale=0.5]{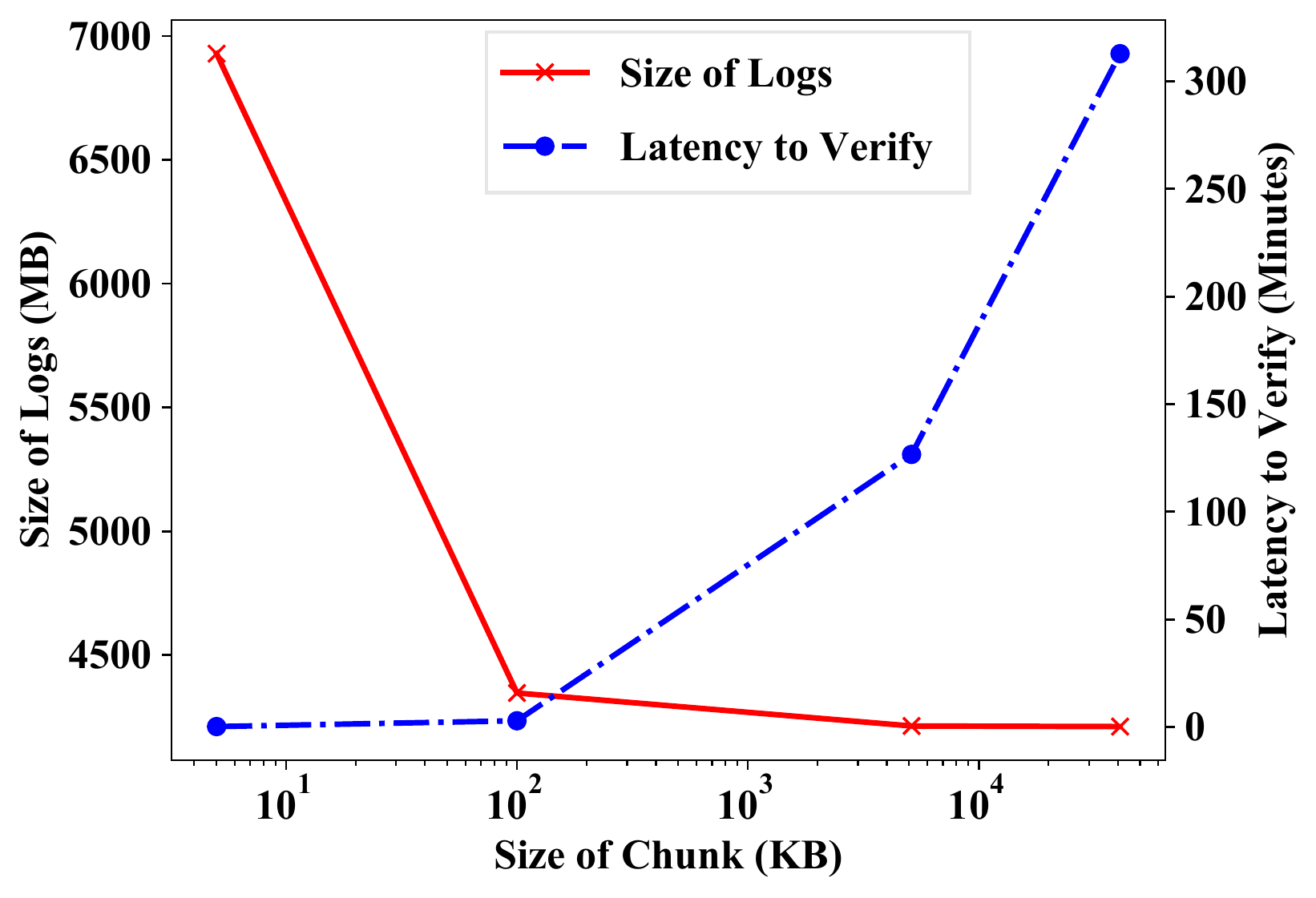}
	\caption{{\color{black} Exp 5: A tradeoff between log size and latency.}}
	\label{fig:latency vs log size}
	\BB
\end{figure}

\begin{figure}[!t]
	\centering
	\includegraphics[scale=0.5]{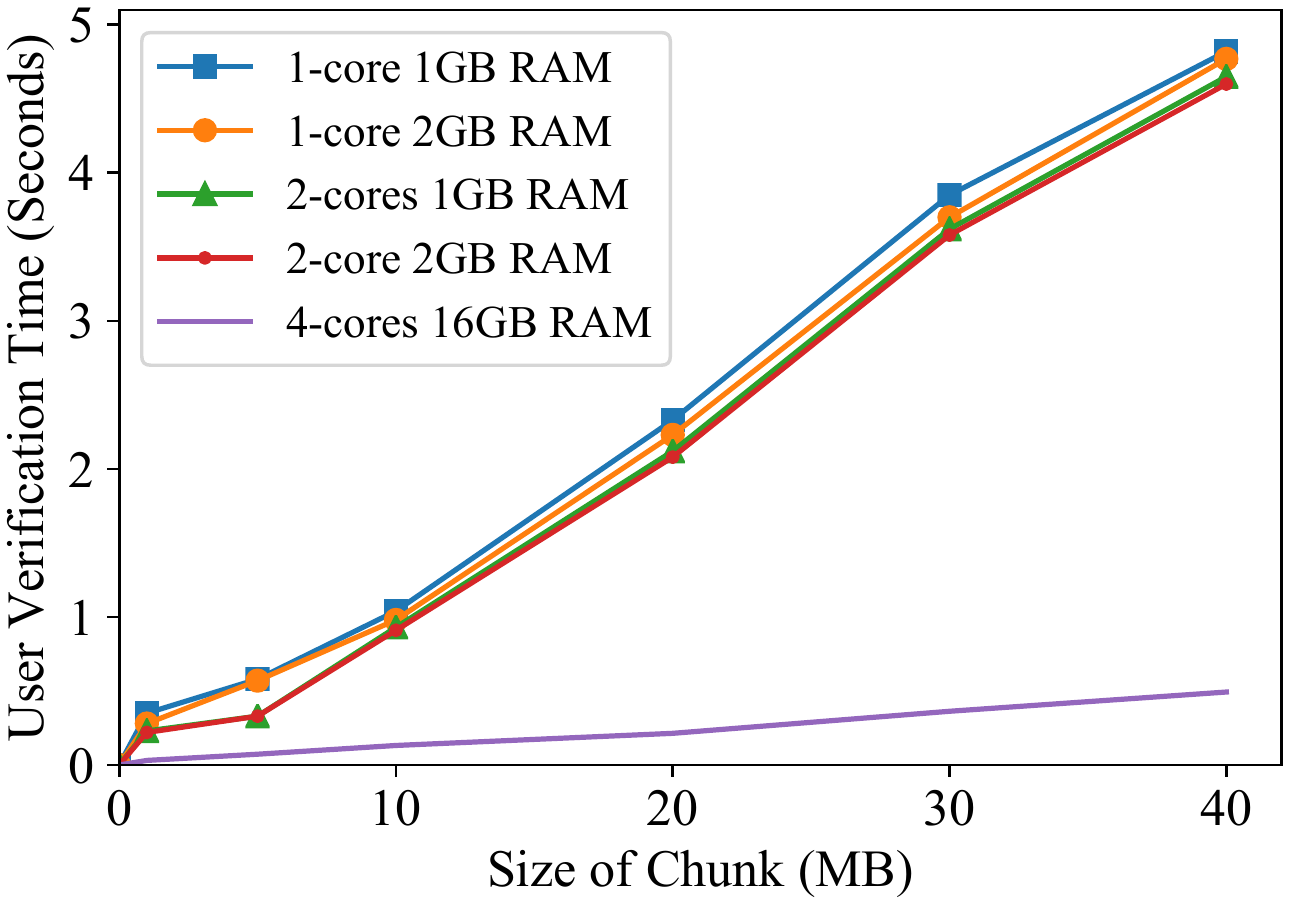}
	\caption{{\color{black} Exp 5: User verification time with different log chunk sizes.}}
	\label{fig:plot-perud-vs-logisze}
	\BB
\end{figure}

\medskip\noindent\textbf{Exp 6: Impact of communication.} We measured the communication impact when a verifier downloaded the sensor data and/or sealed log for attestation. Consider a case when the verifier attests only one-hour/one-day data. The average size of one-hour (one-day) data in a peak hour was 14MB (250MB) having 103K (1.2M) connectivity events, while in a non-peak hour, it was 2MB (50MB) having 13.5K (320K) connectivity events. When using slow (100MB/s), medium (500MB/s), and fast (1GB/s) speed of data transmission, the data transmission time in the case of 1-hour/1-day data was negligible.

\smallskip
\noindent\textbf{Exp 7: Impact of parallelism.} The processing time at each server can be reduced by parallelizing the computation. We investigated the impact of parallel processing to seal a 5MB chunk when having the number of threads 2 or 4 that took 322ms and 310ms, respectively. Increasing more threads did not provide speed-up, since the execution time increases due to thread maintenance overheads. Note that we only parallelized the hash function computation for $\mathcal{PU}$, (while $\mathcal{PI}$ cannot be computed in parallel, due to the formation of hash chains).

\medskip\noindent\textbf{Exp 8. Impact of log optimization.} We compare the impact of per sensor- and per-user-device-id-based optimization methods (denoted by \textsc{Opt-Sensor} and \textsc{Opt-User} in Figure~\ref{fig:plot-perud-vs-logisze}, given in \S\ref{subsubsec:optimization}) against the non-optimized method (denoted by \textsc{Non-Opt} in Figure~\ref{fig:plot-perud-vs-logisze}, given in \S\ref{subsubsec:Sealing Entire Sensor Data}). Since there were 490 WiFi access points in our experiments, we pre-allocated 490 buffers in SGX memory for 490 sensors, one buffer for each sensor. As a result, each buffer size was $\approx$ 85KB.  We also created buffers for groups of devices to implement \textsc{Opt-User}. Here, we again created 490 buffers, and an identical group of user devices is allocated to a buffer. Particularly, the enclave extracted the user device's MAC address, hashed to get the buffer identity, and placed the sensor reading to the corresponding buffer. When the buffer got full, the data in the buffer is cryptographically sealed and written on the disk. Here, we compare storage overhead and verification time.

\begin{figure}[!t]
	\centering
	\includegraphics[scale=0.5]{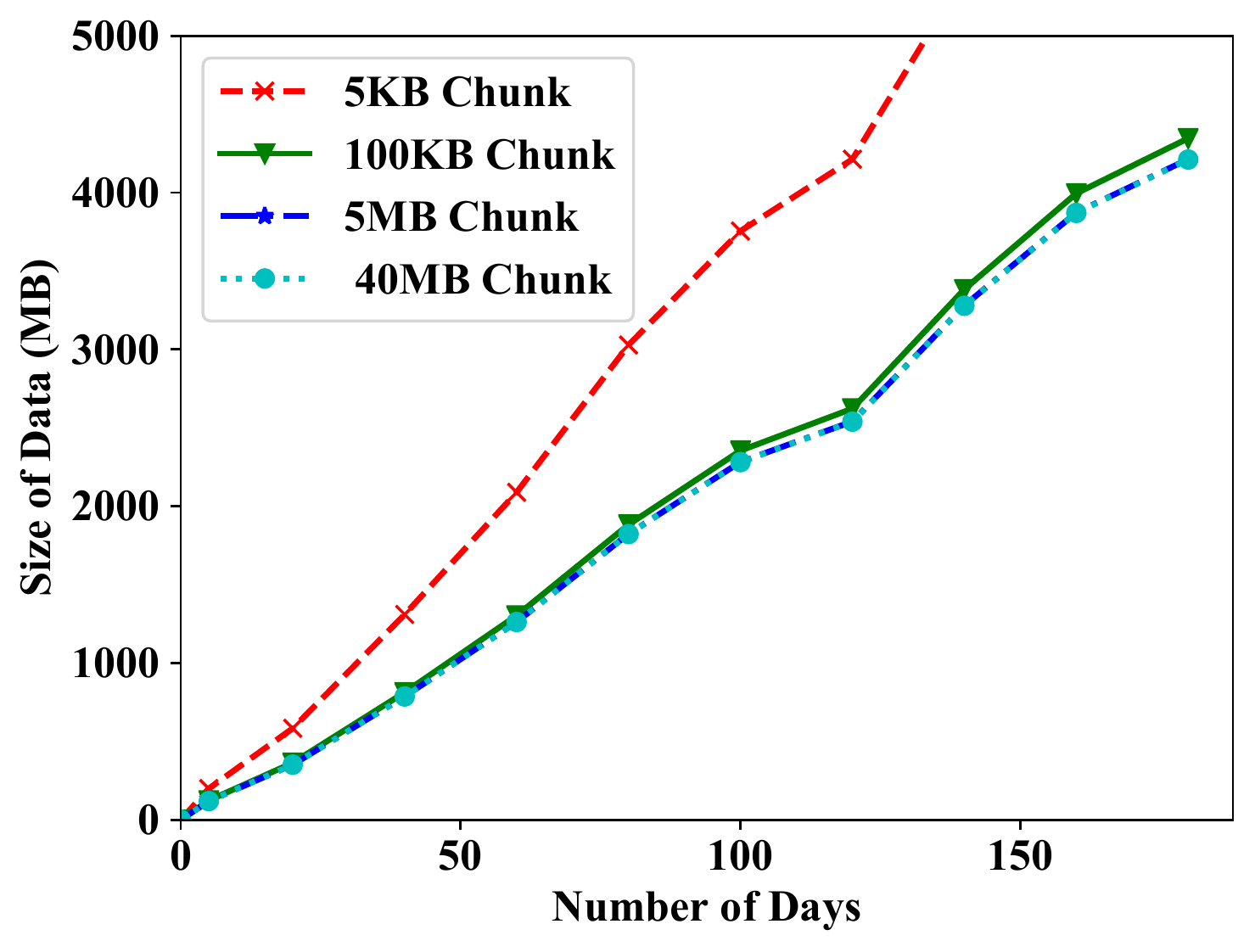}
	\caption{{\color{black} Exp 8: Size of the secured log when using different log sealing methods.}}
	\label{plot-perud-vs-logisze}
	\BBB
\end{figure}

Figure~\ref{fig:plot-perud-vs-logisze} shows the storage overhead to store only cryptographically-secured logs, when using \textsc{Opt-Sensor}, \textsc{Opt-User}, and \textsc{Non-Opt} methods. Since \textsc{Opt-Sensor} and \textsc{Opt-User} methods produce chunks of size 85KB, for a fair comparison, we set the log chunk size to be 85KB in
\textsc{Non-Opt} method. Figure~\ref{fig:plot-perud-vs-logisze} shows that \textsc{Opt-Sensor} and \textsc{Opt-User} save $\approx$6.5\% and $\approx$2.1\% space, respectively, compared to \textsc{Non-Opt} method.

We also investigated the benefit in the performance improvement of user data verification, when using \textsc{Opt-Sensor}, \textsc{Opt-User}, and \textsc{Non-Opt} methods. Note that when using \textsc{Opt-Sensor} and \textsc{Non-Opt} methods, a chunk may store data associated with other users. Thus, it needs to verify additional data, which does not belong to the user. In contrast, when using \textsc{Opt-User} methods, a user has to verify only the desired data that belongs to him/her. It is clear that \textsc{Opt-User} method requires to verify less amount of data, and hence, less verification time, compared to \textsc{Opt-Sensor} and \textsc{Non-Opt} methods. Table~\ref{tbl-perud-verification}  shows verification time and number of chunks required to verify one-day data at
a resource-constrained user (1-core 1GB RAM). Observe that \textsc{Opt-User} method takes significantly less time compared to \textsc{Non-Opt} method.

\begin{table}[!h]
{\color{black}
	\centering
	\begin{tabular}{|l|l|l|}\hline
		Method                & \# chunks & Verification time \\\hline
		\textsc{Non-Opt}      & 3012      & 31.2s             \\\hline
		\textsc{Opt-Sensor}   & 86        & 0.89s             \\\hline
		\textsc{Opt-User}     & 57        & 0.71s             \\\hline
	\end{tabular}
	\caption{{\color{black} Exp 8: User verification performance when using different methods.}}
	\label{tbl-perud-verification}
	\BBB
}
\end{table}
}

{\color{black}

\medskip\noindent\textbf{Summary of the experiments.} We conduct extensive experiments on a real-world setting to evaluate \textsc{IoT Notary}'s overhead, performance, and impact of critical parameters. The results can be interpreted and presented from two principal stakeholders' perspectives in the proposed framework: the service provider and users. On the service provider side, deploying \textsc{IoT Notary} incurs a little overhead. In a default setting, only 21\% of additional storage is needed to retain the tamper-proof logs (Exp 1), and Enforcing policy and generating logs processes for 30-minute data take sub-second computation time (Exp 2). The service provider needs to select the appropriate parameters for an efficient deployment according to its requirements. For example, the chunk size represents the trade-off between the latency for verification and log storage size (Exp 5). Besides, according to different system configurations and data patterns, the service provider could further improve the performance and reduce the overhead by increasing parallelism or incorporating optimized logging algorithms (Exp 7 and Exp 8). On the auditor or users' side, their interactions with \textsc{IoT Notary} is for log verification, and \textsc{IoT Notary}'s verification algorithms allow both the auditor and users to verify data efficiently on different device configurations without incurring computational overhead (Exp 3 and Exp 4).
}

\section{Comparison with Existing Work}
\label{sec:Comparison with Existing Work}

{\color{black} We classify the related work in the scope of IoT attestation into the following four categories:
(\textit{i}) the work that verifies the internal memory of sensors to verify that sensors have not tampered,
(\textit{ii}) the work that verifies the data integrity,
(\textit{iii}) the work that focuses on privacy-preserving data capturing and data release, and
(\textit{iv}) the work that performs auditing in decentralized storage networks.}

{\color{black}
\smallskip
\noindent\textbf{Attestation of the state of the device in IoT.} In IoT settings, several techniques have been developed to verify the internal memory state of untrusted devices through a trusted remote verifier. (Note that such a type of verification is different from the data verification done by \textsc{IoT Notary}.)}  For example, AID~\cite{atone} attests the internal state of neighboring devices through key exchange and proofs-of-non-absence. 
In AID, the adversary can compromise communication channels, the internal state of the device, and the cryptographic keys. SEDA~\cite{atthree} attests low-end embedded devices in a swarm and provides the number of devices that pass attestation. However, SEDA attests neighboring peer devices only. Similarly, DARPA~\cite{attwo} and SANA~\cite{sana16} allow detection of physical attacks by using heartbeat messages and provide aggregate network attestation, with high computation and communication costs, which are quadratic in the network size. SMARM~\cite{atfour} protects against malware by scanning memory in a secret randomized order. However, it may require many iterations to eventually detect malware. RADIS~\cite{radia} assumes that a compromised IoT service can influence genuine operations of other services without changing the software. Thus, RADIS provides control-flow attestation for distributed services among interconnected IoT devices. {\color{black}~\cite{midare} provided SGX-based attestation method for physical attacks on the sensor, \textit{e}.\textit{g}., modifying memory and changing I/O signals.} In short, all such work only deals with attestation of sensor devices, and their methods cannot be used to verify sensor data against the data-capture rules.

In contrast, \textsc{IoT Notary} does not deal with the verification of the internal state of sensor devices, since in our case, (WiFi access-point) sensors are assumed to be deployed by a trusted entity (\textit{e}.\textit{g}., the university IT department). Of course, cyberattacks are possible on sensors to maliciously record data and that can also be detected by \textsc{IoT Notary}, while not verifying the sensors.



\smallskip
\noindent{\color{black}\textbf{Data integrity verification.} To verify data integrity, several protocols based on zero-knowledge proofs, multi-party
computation (MPC), and Merkle trees have been proposed. For example,}
~\cite{chef,kandrian} proposed a privacy-preserving scheme based on zero-knowledge proofs to detect log-exclusion attacks.~\cite{chef} provided solutions for accountability and auditing through hierarchical multi-party computation (MPC) and succinct zero-knowledge proof statements.~\cite{murat08} considered verification process for MPC using a trusted third-verifier.~\cite{kandrian} provided a privacy-preserving certificate transparency service, which signs a message four times, where an auditor can trace the certified logs. Other techniques have proposed the concept of \textit{excerpts} and \textit{snapshots} for log integrity verification. For example, in~\cite{waters2004building} used hash-chains for integrity protection and identity-based searchable encryption.~\cite{treepaper} proposed a Merkle tree-based history tree to prove the sequence of logs over time.~\cite{bloompaper} proposed a Bloom tree that stores proof of logs at an untrusted cloud. Further, access-pattern-hiding cryptographic techniques~\cite{DBLP:conf/eurocrypt/BoyleGI15,DBLP:conf/sp/ZhangGKPP17} may be used to verify any stored log, since an adversary cannot skip the computation on some parts of the data, due to executing an identical computation on each sensor reading. Techniques, \textit{e}.\textit{g}., function secret-sharing~\cite{DBLP:conf/eurocrypt/BoyleGI15} or vSQL~\cite{DBLP:conf/sp/ZhangGKPP17}, may be used to verify query results on cleartext. However, these techniques cannot detect log deletion. Also, all such techniques incur significant time. For example, vSQL took more than 4000 seconds to verify a SQL query. In addition, any end-to-end encryption model is not sufficient for verification.

In contrast, \textsc{IoT Notary} provides complete security to sensor data and a real-time data attestation approach. Unlike~\cite{kandrian}, \textsc{IoT Notary} requires only two signatures per file, where one is used to validate log completeness, and another is used to validate the log ordering with respect to adjacent logs.

{\color{black}
\smallskip
\noindent\textbf{IoT data privacy policy enforcement.} Several works propose solutions for data privacy policy enforcement before the data is released for any service provisioning. From an architectural perspective, these solutions could be coarsely classified into two categories: \emph{middle-box methods} and \emph{on-device methods}, depending on where the policy enforcement functions are placed. The \emph{middle-box methods} import a new trusted and commonly centralized service that aggregates and processes the data according to the privacy policies before the data is released to any service provider. Personal data stores (PDS)~\cite{pdv,openpds} provide a general rule-based access management framework for sensitive personal data, including mobile sensor data. Besides access control functions, the model that ~\cite{databox} introduced also helps the data controller and service providers demonstrating compliance with privacy regulations. ~\cite{privacy-mediator} combines policy enforcement with edge computing and proposes a solution where the policy enforcement functions could be deployed flexibly at a location in a user's trusted domain. The middle-box methods have been adopted for processing health data~\cite{medical1,medical2,medical3}, location data~\cite{beresford2004mix, location_privacy}, and video data~\cite{video-privacy-cloudlet}.

In contrast to introducing an additional trusted entity in the system, the \emph{on-device methods} aim to enforce the privacy policies at the data source directly. Achieving this goal requires more computational power on the sensing device. Thus, on-device methods are generally designed for a mobile computing setting, where a smartphone works as a sensing device. ~\cite{ipsheild} and~\cite{semadroid} propose general frameworks for privacy and context-aware sensor management on smartphones for users to define their sensor data access policy in a fine-grain manner. In addition, the pervasive photograph-capturing capability on mobile devices causes privacy concerns nowadays~\cite{camera_concerns}. Several solutions have been proposed to ensure data capturing policies for video feeds on mobile devices, including smartphones~\cite{cardea,ipic}, wearable devices~\cite{wearable}, and augmented reality (AR) devices~\cite{world-driven}.

Unlike existing middle-box solutions, \textsc{IoT Notary} minimizes infrastructure deployer (IFD)'s overhead. It does not require the IFD to implement the data capturing policy enforcement and logging functions in its existing infrastructure. Instead, the service provider (SP) who eventually uses the data will pay overhead. In our experience, this model is critical for the agreement of collecting data from IFDs, because infrastructure and organizations, \textit{e}.\textit{g}., network infrastructure, water infrastructure, and building management, may not have extra resources for additional services other than their regular workload. Additionally, \textsc{IoT Notary}'s model also leads to better scalability of supporting more SPs. When a new SP requires data from IFD, IFD only needs to spawn the trusted agent program for the new SP initially. In operation, the communication overhead of sending the encrypted data to the new SP is the only additional overhead. On the other hand, comparing with on-device solutions, \textsc{IoT Notary} supports existing commodity devices without modifications. This advantage is vital for supporting heterogeneous devices in an IoT scenario.  

}

{\color{black}
\smallskip
\noindent\textbf{Auditing in decentralized storage networks.}
Decentralized storage networks (DSN) refer to a group of untrusted nodes (or people) at multiple locations who are incentivized to join the network, store the data of other users, and keep data accessible. Since the data is stored at multiple untrusted nodes in DSN, several data auditing schemes such as centralized private auditing, Merkle tree-based auditing, succinct proof
framework, SNARK frameworks, data tagging, and Pedersen commitment-based zkLedger ~\cite{DBLP:conf/nsdi/NarulaVV18,filecoin,du2021enabling,vorick2014sia,DBLP:conf/crypto/Ben-SassonCGTV13,DBLP:conf/sp/ParnoHG013} have been proposed. In contrast, \textsc{IoT Notary} provides a central data repository and mechanisms for data attestation by several users. }

\section{Conclusion}
\label{sec:Conclusion}
This paper presented a framework, \textsc{IoT Notary} for sensor data attestation that embodies cryptographically enforced log-sealing mechanisms to produce immutable proofs, used for log verification. Our solution improves the na\"{\i}ve end-to-end encryption model, where retroactive verification is not provable. The service verification mechanism on failing at users allows them to revoke services of the concerned IoT space. Therefore, a user is not required to blindly trust in the IoT space, and we empower the users with the right-to-audit instead of right-to-own the data captured by sensors. \textsc{IoT Notary} is a part of a real IoT system (TIPPERS) and provides verification on live WiFi data with almost no overheads on users.

{\color{black}At the high level, \textsc{IoT Notary} works in three phases, as follows: (\textit{i}) Notification, in which the system informs the users about the data capturing policies established by the organizations;
(\textit{ii}) Log sealing, in which the system stores the received sensor data in a secure manner using hash chains and produces the proof; and (\textit{iii}) Verification, in which the user verifies the presence/absence of their data against the data capturing policies.
We, future, provided two techniques to optimize the secure logging sealing process, by creating per sensor-based logging and per user-based logging. The experimental results show the effectiveness of \textsc{IoT Notary} in a real IoT environment, consisting of 490 WiFi access-points across 30 buildings of the University of California, Irvine. The experiments show that the overheads of \textsc{IoT Notary} in creating the secure log, storing the log, as well, in verifying secure logs at the resource-constrained users are affordable.}

\noindent\textbf{Future directions.}
 Now, we provide the future directions that one can explore and improve \textsc{IoT Notary}, as follows:
\begin{enumerate}
{\color{black}

    \item \textbf{Contextual policies.} While we presented \textsc{IoT Notary} for the case of monotonous data capturing policies, which are applicable to all the users, an interesting direction would be in exploring the contextual data capturing policies, where the context is determined based on data captured by other sensors. Extending the system to handle such data-capture rules, \textit{e}.\textit{g}., ``Do not capture my WiFi connectivity data if I am the only person connected to the access point,'' is a non-trivial challenge.

    \item \textbf{User registration.} \textsc{IoT Notary} currently works in an organization that maintains a user-device registry for their authentication. However, such a registry may not exist in spaces such as shopping malls or stadiums. A potential improvement in \textsc{IoT Notary} addressing this issue could be authorizing verification query based on device identity and only allowing a device to verify its own data. However, ensuring the integrity of the device identifier remains a challenge.

    \item \textbf{Device spoofing.} The current \textsc{IoT Notary} implementation for wireless network data utilizes the MAC address as the device identifier to enforce data capturing policies. However, the method may be vulnerable to identifier spoofing attacks. It is important to design or utilize an appropriate device authentication mechanism in the network and use more secure device identifiers.

    \item \textbf{Sensor spoofing.} \textsc{IoT Notary}'s enforcement of data-capturing rules relies on the assumption that the service provider cannot tamper with the deployment of sensors. For example, swapping sensors in two rooms may cause the data capturing inconsistent with the published rules. Thus, another interesting direction of work would be in enhancing \textsc{IoT Notary} for ensuring trust on sensor deployment. Potential solutions, including building sensor registry and verifying sensors at the network level, pose new challenges.

    \item \textbf{A very-large-scale deployment.} The current \textsc{IoT Notary} system works in a campus-scale deployment, where hundreds to thousands of sensors exist. It is interesting to explore how the system could be extensively deployed in a city-scale scenario, where a massive amount of sensors generate data. It will require \textsc{IoT Notary} to {processing data obtained from multiple trusted agents, in parallel}. This improvement proposes new challenges like load balancing and data-capturing rules consistency at a wide coverage area.
}
{\color{black}
\item \textbf{Extending \textsc{IoT Notary} for decentralized storage networks.} The current \textsc{IoT Notary} system works in a centralized storage system, and extending \textsc{IoT Notary} for DSN comes with challenges such as: how to inform users where their data has been stored, how the users will communicate with different nodes in DSN, what will happen if data is stored according to per sensor or per user, and what will be the overhead of using DSN compared to the centralized system.}
\end{enumerate}

\end{document}